\documentclass[twocolumn]{aastex631}
\usepackage{amsmath}	
\usepackage{amssymb}	
\usepackage{multirow}
\usepackage{CJK}

\newcommand{\Mg}{M_{\mathrm{g}}}
\newcommand{\Msc}{M_{\mathrm{sc}}}
\newcommand{\Rg}{R_{\mathrm{g}}}
\newcommand{\rt}{r_{\mathrm{t}}}
\newcommand{\rtz}{r_{\mathrm{t0}}}
\newcommand{\Rd}{R_{\mathrm{d}}}
\newcommand{\Rs}{\overline{R}}
\newcommand{\rh}{r_{\mathrm{h}}}
\newcommand{\rl}{r_{\mathrm{lagr}}}
\newcommand{\rc}{r_{\mathrm{c}}}
\newcommand{\rcut}{r_{\mathrm{cut}}}
\newcommand{\trh}{T_{\mathrm{rh}}}
\newcommand{\trhz}{T_{\mathrm{rh0}}}
\newcommand{\tcr}{T_{\mathrm{cr}}}
\newcommand{\tcrz}{T_{\mathrm{cr0}}}
\newcommand{\fbh}{f_{\bullet}}
\newcommand{\Mbh}{M_{\bullet}}
\newcommand{\rhog}{\rho_{\mathrm{g}}}
\newcommand{\rhogz}{\rho_{\mathrm{g0}}}
\newcommand{\rhosc}{\rho_{\mathrm{sc}}}

\newcommand{\amax}{\alpha_{\mathrm{max}}}
\newcommand{\Omegag}{\Omega_{\mathrm{g}}}
\newcommand{\Omegabh}{\Omega_{\bullet}}
\newcommand{\gammag}{\gamma_{\mathrm{g}}}
\newcommand{\gammabh}{\gamma_{\bullet}}
\newcommand{\rrho}{\mathcal{R_{\mathrm{\rho}}}}

\shorttitle{Star cluster }
\shortauthors{Wang et al.}
\graphicspath{{./}{figures/}}

\begin{document}
\begin{CJK*}{UTF8}{gbsn}
  \title{Tidal disruption of stellar clusters and their remnants' spatial distribution 
    near the galactic center \footnote{Released on May, 8th, 2022}}

  \author[0000-0001-8713-0366]{Long Wang (王龙)}
  \affiliation{School of Physics and Astronomy, Sun Yat-sen University, Daxue Road, Zhuhai, 519082, China}
  \affiliation{CSST Science Center for the Guangdong-Hong Kong-Macau Greater Bay Area, Zhuhai, 519082, China}

  \author[0000-0001-5466-4628]{D. N. C. Lin (林潮)}
  \affiliation{Department of Astronomy and Astrophysics, University of California, Santa Cruz, CA, USA}
  \affiliation{Institute for Advanced Studies, Tsinghua University, Beijing 100086, People’s Republic of China}



  \begin{abstract}

    The accretion of massive star clusters via dynamical friction 
    has previously been established to be a likely scenario for the build 
    up of nuclear stellar clusters (NSCs). 
    A remaining issue is whether strong external tidal 
    perturbation may lead to the severe disruption of
    loosely-bound clusters well before they sink deeply into the 
    center of their host galaxies.  
    We carry out
    a series of $N$-body simulations and verify our early idealized
    analytic models. We show if the density profile of the host galaxies can be described by 
    a power-law distribution with an index, $\alpha <1$, 
    the cluster would be compressed in the radial direction by the external galactic tidal field. 
    In contrast, the galactic tidal perturbation is 
    disruptive in regions with a steep, $\alpha >1$, density fall-off or in the very center where 
    gravity is dominated by the point-mass potential of super-massive black holes  (SMBHs). 
    This sufficient criterion supplements the conventional necessary 
    Roche-lobe-filling condition in determining the preservation versus disintegration 
    of satellite stellar systems.
    We simulate the disruption of 
    stellar clusters which venture on nearly-circular, modestly- or highly-eccentric orbits into 
    the center of galaxies with a range of background density profiles and SMBHs. We 
    obtain the spatial distribution of the stellar-cluster remnants.
    We apply these results to the NSC within a few parsecs 
    from SMBH Sgr A$^\ast$ at the Galactic Center.  Recent observations indicate the
    coexistence of two populations of stars with distinctively separate
    ages and metallicities. 
    We verify that the subsolar-metalicity population can be 
    the debris of disrupted stellar clusters.
  \end{abstract}



  \section{Introduction} 
  \label{sec:intro}
  The widely adopted $\Lambda$CDM model of galaxy formation is based on the assumption 
  that relative dense early-generation dwarf galaxies form, merge, and assemble into
  larger entities\citep{white1978, blumenthal1984}.  A prediction of this hypothesis 
  is the preservation of ubiquitous satellites which survived the tidal disruption 
  during their dynamical evolution\citep{navarro1995}.  Recent discoveries of many 
  debris stellar streams provide supporting evidences for this scenario
  \citep{myeong2018, helmi2018}.  If their dense nuclei with sufficient mass
  ($\gtrsim 3 \times 10^7 M_\odot$) can be preserved, they 
  may converge towards the centers of amalgamated stellar systems and
  their merged halo\citep{fall1977}.

  Nuclear stellar clusters are also commonly found in galaxies 
  (see the extensive contributions by many investigators cited in two annual review articles
  by \citet{kormendy2013} and \citet{Neumayer2020}).
  A natural extrapolation is that these clusters were formed in the inner ($<$ a few kpc) region 
  and migrated to the center of their host galaxies under the action of dynamical friction
  \citep{tremaine1975, tremaine1976}.  Near the central region of the Galaxy, there are several stellar
  clusters including Archies and Quintuplet \citep{nagata1995, cotera1996}.
  Within a few pc from the Sgr A$^\ast$ supermassive black hole (SMBH) \citep{genzel1997, ghez1998}, 
  there is a nuclear cluster with $\sim 1-2 \times 10^7$ mature stars \citep{do2009, schodel2014} 
  in addition to the $\sim 10^2$ bright young massive S and disk stars \citep{ghez2003}.
  The nuclear-cluster stars show substructure in kinematics, heavy element abundance, 
  and stellar ages \citep{feldmeier2014,do2020}. 

  In addition to the Occam's-razor {\it in-situ} formation scenario for the origin of nuclear 
  clusters\citep{loose1982, agarwal2011}, it has been widely suggested that they
  contain stars formed in progenitor clusters beyond a few Kpc, endured orbital
  decay under the action of dynamical friction \citep[e.g.,][]{Capuzzo-Dolcetta1993, oh2000,
    Lotz2001,Capuzzo-Dolcetta2009, Antonini2013, feldmeier2014,Gnedin2014,Arca-Sedda2014}.
  Several series of N-body simulations provided quantitative supports for this hypothesis \citep[e.g.,][]{oh2000, Capuzzo-Dolcetta2008,Antonini2012,Perets2014,Arca-Sedda2015,
    Arca-Sedda2016,Arca-Sedda2017a,Arca-Sedda2017b,arcasedda2018,tsatsi2017,arcasedda2020}.
  They have already demonstrated this scenario can account for the origin of common-rotation 
  and diverse-abundance properties among subgroups of nuclear cluster stars
  \citep[e.g.,][]{feldmeier2014,tsatsi2017,Fahrion2020, arcasedda2020}. 

  Nevertheless, there is a remaining issue of how close to the Galactic center 
  can the clusters deliver a substantial fraction of their constituent stars.  
  Loosely bond stellar clusters and satellite dwarf galaxies have a tendency to undergo
  tidal disruptions within a conventionally-defined ``tidal disruption distance'' $R_{\rm d}$ at 
  Galactocentric distances of a fraction to a few kpc \citep{fall1977, oh1992, 
    oh1995, fellhauer2007}. The central objective of this paper is to examine this 
  ongoing dispersal issue based on the assumption that the parent clusters can make 
  their way towards the Galactic center.  

  The Galactic potential has a complex radial
  dependence which has been approximated as a composite of several components, including the
  bulge, disk, halo \citep{Gnedin2005, Widrow2005}, and the SMBH.
  Moreover satellite galaxies and stellar clusters have diverse internal  
  structure\citep{Baumgardt2019,Baumgardt2021}.  
  Many investigators have studied how star clusters can survive in the effect of tidal 
  disruption during their arduous journey through different regions and at various 
  evolutionary stages of their host galaxies.
  In some cases, they combine semi-analytic treatments to model star cluster dynamics with actual 
  galaxy models from magneto hydro-dynamical cosmological simulations\cite[e.g.][]{Gnedin2014, 
    Longmore2014, Kruijssen2014, Kruijssen2015, Pfeffer2018, Choksi2018, Li2018}.  Near the 
  Galactic center, if the mass of SMBH significantly exceeds the mass of infalling GCs, they 
  would be completely disrupted before they reach a few pc from the SMBH \citep{Arca-Sedda2015}. 
  This issue is directly relevant to the dynamical structure at the very center of NSCs and 
  perhaps the formation of SMBHs in galaxies.   Fittings of the observed surface brightness 
  with the \citet{sersic1968} model show a wide variation in the light and mass distribution 
  among different galaxy populations, including those with or without nuclear clusters 
  \citep{boker2002, misgeld2011, kormendy2013}. The poorly-resolved surface-brightness 
  and the inferred mass-density distribution within the central few pc from any SMBH 
  at the galactic centers also vary considerably.  A complementary study on the disruption 
  and survivability of stellar clusters in a diverse set of galactic potential is warranted.

  Motivated by this generic problem, we carried out an investigation on the secular evolution 
  of stellar clusters in a general galactic tidal field \citep{Ivanov2020}.  Based on a 
  rigorously-constructed, idealized formalism \citep{Mitchell2007}, we analytically 
  showed that the tidal field of a background potential associated with a sufficiently 
  flat density distribution can lead to compression rather than disruption.  We also 
  showed that even the most vulnerable homogeneous star cluster (in contrast with the 
  more tightly-bound models with dense cores or intermediate-mass black holes, IMBHs) can 
  meander to distances much smaller than the conventional $R_{\rm d}$ without disruption, 
  thus potentially contribute to the accumulation of stars in the NSCs.  Although these 
  analytic approximations provide a quantitative illustration on the critical conditions, 
  they are derived for clusters with idealized, uniform density distribution. In order to 
  verify and generalize the results of our previous analytic approximation,  we carry out 
  a series of numerical N-body simulations with more a general  \cite{michie1963}-\cite{king1966} 
  model for the  initial stellar density distributions.  In \S\ref{sec:nbody}, we briefly describe 
  the numerical method, initial and boundary condition. In  \S\ref{sec:circular}, we simulate
  the evolution of a cluster's internal density distribution.  For initial conditions, we
  chose the cluster's density to be homogeneous or follows King models with various concentration
  parameter.  The cluster is initially on a circular orbit around galactic potential associated
  with various mass distribution, including a point mass to represent a SMBH.  
  We consider, in \S\ref{sec:evolve} the survival of clusters with gradually
  decaying (nearly circular) and plunging (highly eccentric) orbits.
  In \S\ref{sec:summary}, we summarize the results of our numerical simulation and discuss their 
  implications including the possibilities that in-situ star formation may also be a major 
  contribution to the NSC build-up \citep[e.g.][]{Antonini2015, portegies2002} and the Milky-Way 
  NSC may be built-up by both accreted massive star clusters and in-situ star formation, as shown 
  from Chemo-dynamical analysis by \cite{do2020}.

  \section{Numerical simulations}
  \label{sec:nbody}

  We aim to investigate whether a star cluster can  survive the tidal 
  perturbation induced by the galactic potential. If the cluster 
  is severely or completely disrupted during its passage through the 
  conventional tidal-disruption radius $\Rd$, its tidal debris would form a ring 
  along that radius and there would not be any subsequent mechanism that can 
  efficiently bring the detached stars closer to the galactic center.  But if the 
  cluster is tidally compressed, it would survive and migrate well inside $\Rd$, 
  as suggested by \cite{Ivanov2020}. To verify whether debris stars can reach 
  the SMBH proximity, we carry out a series of N-body simulations, to demonstrate 
  the effects of tidal compression and disruption, depending on the density profiles of 
  the star clusters, the potentials of the galaxies, and the orbits of the clusters.

  \subsection{N-body code}

  In this work, we use the $N$-body code \textsc{petar} \citep{Wang2020b} to perform the numerical simulations of the star clusters. 
  The code is designed for simulating dense stellar systems where close encounters and dynamics of binaries are important. 
  The particle-tree and particle-particle methods \citep{Oshino2011}, embedded in the framework for developing parallel particle 
  simulation codes (\textsc{fdps}), are used to achieve a high computing performance \citep{Iwasawa2016,Iwasawa2020}.
  The slow-down algorithmic regularization method \citep[SDAR;][]{Wang2020a} is designed and implemented to
  accurately follow the orbital motions of binaries, hyperbolic encounters and hierarchical few-body systems.
  The tidal force from the Galactic potential is calculated with the \textsc{galpy} code \citep{Bovy2015}. 

  \subsection{Star cluster model}

  In our $N$-body simulation, we neglect stellar evolution and non-uniform mass function,
  i.e. all stars are assigned with the same mass and lifespan longer than the
  computational time span.  In principle, the stellar-wind mass loss and 
  the phase-space segregation of multiple-mass components can affect the 
  dynamical evolution of star clusters and subsequently influence the 
  survival of star clusters \citep{portegies2002}.  In general, these physical 
  processes cannot be ignored. But the mixture of them in one set of simulations 
  would introduce some difficulties in disentangling their relative impacts.
  For the purpose of this investigation, we adopt an idealized approximation to keep 
  the $N$-body models relatively simple.

  The total initial number of stars in most $N$-body models are fixed to be 1000. 
  At the beginning of each simulation, the system is constructed to be in a virial 
  equilibrium.  This model does not fully represent a genuine globular cluster
  in nature, which typically contains million evolving stars with a range of masses. 
  But it is time-consuming to carry out such comprehensive simulations.  In this
  investigation, we simulate many models with a wide range of other parameters, including
  the clusters' internal and galaxies' external mass distribution as well as 
  clusters' orbital evolution (Table \ref{tab:init}). With limited computational 
  resources, it is practical
  to simplify and speed up the simulations with cluster models more sensitive to the 
  tidal effect and neglect less-dominant relaxation effects.  Nevertheless, we compare
  the results of tidal-response models with NoTide calibration models and verify 
  that our results are not significantly affected by spurious internal two-body 
  relaxation effects during the simulated time intervals.  Our analysis does not lose 
  generality, since the external (Galactic) tidal effect is always present, regardless 
  of the clusters' mass, albeit their evolutionary time scale may vary.

  We adopt two types of initial-density profile for the clusters: an idealized spherical-symmetric,
  homogeneous mass-density distribution\cite{Mitchell2007} and a series of \cite{michie1963}-
  \cite{king1966} models. The former setup is the simplest model where the tidal effect can 
  be described and analyzed with an analytic approach. These clusters are also most vulnerable
  to external perturbation.  We adopt it to validate the prediction of \cite{Ivanov2020}.
  But this idealized homogeneous model is unrealistic with respect to the observed 
  star clusters. Thus, we carry out additional simulations with the 
  \cite{michie1963}-\cite{king1966} profile, which is commonly adopted to describe the 
  observed surface-brightness distribution of globular clusters with a dense core and 
  tidal cutoff of the outer region.

  Most of our models contain $10^3$ stars which is well below the star counts
  inside typical globular clusters in nature.  In order to build-up statistical
  significance with such small number of cluster stars, we usually need to carry out 
  many simulations with the same initial condition but different random seeds 
  to generate the positions and velocities of stars. We obtain some average trends 
  among these models to represent the mean expectation values and to smooth out 
  any stochastic scatter.  In this work, we only prepare one initial model 
  (one random seed) for each type of density profile.  With this approach, when 
  we choose a certain density profile and compare the effect from different Galactic 
  potentials, we are ensured to compare the simulations with the identical initial 
  positions and velocities of stars.  This prescription can also reduce the impact 
  from small-number stochastic scatter.  With a series of statistical tests 
  (Fig. \ref{fig:randcheck}), we verify that 
  this approach is sufficient in our analysis, since the tidal effect is very pronounced.

  \subsubsection{Units and scaling}

  To describe the dynamical evolution of star clusters, we calculate two important timescales: 
  the crossing time ($\tcr$) and the two-body relaxation time ($\trh$).  The crossing time
  $\tcr$ is defined to be
  \begin{equation}
    \tcr \equiv \sqrt{\frac{\rh^3}{G \Msc}},
    \label{eq:crossingtime}
  \end{equation}
  where $\rh$ is the half-mass radius of a cluster.
  For equal-mass system, $\trh$ can be described as \citep{Spitzer1987}
  \begin{equation}
    \trh \approx 0.138 \frac{N^{1/2} \rh^{3/2}}{m^{1/2} G^{1/2} \ln \Lambda},
    \label{eq:trh}
  \end{equation}
  where $\Lambda=0.4 N$ and $m$ is the mass of star.
  We use the initial $\trh$ ($\trhz$) as the time unit and the cluster's initial cut-off (tidal) radius $\rt$ (Eq. \ref{eq:rt}) as the radial unit.
  Thus, the result can be scaled to arbitrary systems with a free choice of $\Msc$ and $\rh$, and it can also represent star clusters with different $N$ but the same $\trh$.

  \subsubsection{Idealized clusters with spherically symmetric homogeneous density profile}
  \label{sec:homogenous}
  For a star cluster moving in a circular orbit, the conventional tidal radius of the cluster can be approximated as
  \begin{equation}
    \rt(R) \approx \left[ \frac{\Msc}{3\Mg(R)} \right]^{\frac{1}{3}} R
    \label{eq:rt}
  \end{equation}
  where $\Msc$ is the total mass of the cluster.
  The density profile of a homogeneous star cluster can be described as
  \begin{equation}
    \rhosc(r) = 
    \begin{cases}
      \frac{4 \Msc}{3 \pi r_{\rm cut}^3} &  \text{for}~r < \rcut \\
      0   &     \text{otherwise} 
    \end{cases} 
    \label{eq:rcut}
  \end{equation}
  where $r$ is the distance to the center of a cluster (\S\ref{sec:homo}), and $\rcut$ is the cut-off radius of the cluster.
  The velocity distribution of stars follows the Maxwell distribution with a distance dependent scaling factor:
  \begin{equation}
    f(v,r) = \sqrt{\frac{2 G \Msc}{3 \pi}} v^2 \exp{\left(\frac{-v^2}{2} \right)} \frac{r}{\rcut}.
  \end{equation}
  We place the clusters at $R=\Rg$ from the center of the galactic potential where $\Rg$ is a reference distance 
  (also see Eq. \ref{eq:ps}) and the clusters' $\rcut$ is set to be $\rt$.  This location corresponds to the conventional
  galacto-centric distance where the clusters are assumed to be on the verge of tidal disruption.

  \subsubsection{The Michie-King model of stellar clusters}
  \label{sec:michie}

  The \cite{michie1963}-\cite{king1966} model describes a stellar system with a non-singular isothermal sphere.
  There are two free parameters: $\rcut$ and the concentration parameter $W_0$, which indicates the 
  ratio between $\rcut$ and the core radius ($\rc$). In observational interpretation, $\rcut$ is often considered to be $\rt$ .
  We also adopt this convention in most models presented here such that $\Rg$ is the conventional
  galacto-centric distance where the clusters are assumed to be on the verge of tidal disruption and {\it the conventional, necessary, Roche-lobe-filling condition for tidal disruption is $R \leq \Rg$.}
  To verify whether the computational results are independent of the initial conditions (\S\ref{sec:twocomponents}),
  $\rt$ and $r_{\rm cut}$ are chosen separately in some test models.
  A part of our models adopt three Michie-King profiles with $W_0=2,6$ and $8$ (hereafter named as W2, W6 
  and W8), respectively (\S\ref{sec:kingw0}).
  The W2 profile has a low concentration, the W6 profile is similar to the \cite{Plummer1911} profile.
  Due to the low number of stars, the W8 profile does not show significant difference of central density referring to that of W6.
  Thus, these three are sufficient to represent a wide range of density distribution.

  \subsection{Galactic potential}

  \cite{Ivanov2020} suggest that for a static power-law spherically symmetric galactic potential with a 
  small power index ($\alpha$), a star cluster with the homogeneous density profile can suffer tidal 
  compression instead of tidal disruption. 
  Consequently, the cluster can migrate to the inner region of the galactic center.
  Firstly, we carry out $N$-body simulations with this one-component static potential for different values of $\alpha$, 
  in order to validate the theoretical prediction.
  This potential represents that of the galactic bulge.
  We place star clusters with different density profiles on a circular orbit to investigate their morphological evolution.
  We determine the critical value of $\alpha$ which represents the boundary between tidal disruption and compression.

  For the next step, we introduce a point-mass potential superimposed onto the power-law potential to represent 
  galaxies with central SMBHs. 
  We choose the power-law potential with $\alpha=0.5$ which, in the absence of the SMBH, provides the effect of 
  tidal compression.  The additional SMBH provides the counter-effect of tidal disruption.
  We vary the mass ratio between these two components and find the boundary between
  these competing effects.

  Previous investigations have provided well-established evidences that clusters migrate inwards 
  under the action of dynamical friction.  In their presence at the center of the galaxy, 
  SMBHs' tidal influence intensify as  a stellar cluster undergoes orbital decay towards the galactic 
  center.  To reproduce the migration of clusters, we artificially increase the total mass of the potential 
  instead of implementing dynamical friction in the $N$-body code.  In this prescription, 
  the cluster smoothly sinks into the center of the galaxy with an in-spiraling orbit (\S\ref{sec:appendixa}).
  We investigate whether the tidal compression can help the cluster to survive inside the conventional $\Rd$.  
  We also investigate the survival of star clusters with modestly and highly eccentric orbits, 
  where they approach the galactic center at their perigee much faster than any in-spiraling migration
  due to dynamical friction.

  \subsubsection{One-component static background potential}
  \label{sec:onecomponent}

  By defining $R$ as the distance to the galactic center, the density profile for the static power-law 
  spherically-symmetric potential can be written as
  \begin{equation}
    \rhog(R) = \rhogz \left (\frac{\Rg}{R} \right)^\alpha,
    \label{eq:ps}
  \end{equation}
  where $0<\alpha\le 3$, $\Rg$ is a reference distance, and $\rhogz$ is the reference density defined at $R=\Rg$.
  The corresponding enclosed mass of the galaxy at $R$ is 
  \begin{equation}
    \Mg(R) = 4 \pi \rhogz \Rg^\alpha
    \begin{cases}
      \frac{R^{3-\alpha}}{3-\alpha} & \text{for}~\alpha\ne 3\\
      \ln (R) & \text{for}~\alpha=3.
    \end{cases}
    \label{eq:mgr}
  \end{equation}
  The corresponding galactic potential has the form 
  \begin{equation}
    \Psi_{\mathrm{g}}(R) =  4 \pi G \rhogz \Rg^\alpha 
    \begin{cases}
      \frac{R^{2-\alpha}}{(3-\alpha)(2-\alpha)}  & \text{for}~\alpha\ne 2,~\alpha \ne 3 \\
      \ln(R) & \text{for}~\alpha=2  \\
      - \frac{\ln(R)+1}{R} & \text{for}~\alpha=3,
    \end{cases}
  \end{equation}
  where $G$ is gravitational constant.

  For the power-law potential of Equation~\ref{eq:ps}, \cite{Ivanov2020} derives 
  a galacto-centric transitional distance  from tidal compression to
  disruption distance
  \begin{equation}
    \Rd = \left (\frac{3 \alpha^4}{3-\alpha} \right)^{1/\alpha} \Rg
    \label{eq:rtde}
  \end{equation}
  which differs from the conventional tidal disruption radius derived for $R$ from Equation (\ref{eq:rt}) with $\rt=r_{\rm cut}$ ($R \le \Rg$). 
  Clusters with $\Rg \geq \Rd$ are outside the tidal disruption region and they endure tidal compression rather than disruption.
  {\it The sufficient criterion for disruptive tidal perturbation is $R \leq \Rd$}, instead.
  The corresponding tidal-compression criterion is $\alpha \leq \amax \approx 0.913$.
  In order to validate this conjecture, we carry out 5 sets of $N$-body simulations of the star clusters 
  which include three values of $\alpha$ (0.5, 1 and 2) for power-law potentials, a point-mass potential,
  and in the absence of galactic potential, respectively (\S\ref{sec:circular}).
  The corresponding values of 
  $\Rd = 0.005625$, $1.500$ and $6.928~\Rg$ for $\alpha=0.5$, 1 and 2, respectively. The small value 
  of $\Rd/\Rg$ for small $\alpha$ again shows that clusters can be tidally compressed at galacto-centric 
  distance well inside the conventional tidal disruption radius.
  Without the loss of generality, we place these clusters at $\Rg = 100~\rt$ and let them move on a circular orbit.

  For each value of $\alpha$, we perform four sets of simulations with a homogeneous density profile. 
  The results of these simulations are shown in \S\ref{sec:homo} and in \S\ref{sec:kingw0}.
  Using the Michie-King prescription, we focus on the W2 clusters and vary the setups of galactic potentials in the following sections.
  With a low central concentration, these clusters are sensitive to the tidal effect.

  \subsubsection{SMBH's contribution}
  \label{sec:twocomponents}

  The potential of a SMBH can be described by a point-mass potential with the form as
  \begin{equation}
    \Psi_{\mathrm{bh}}(R) =  - \frac{G \Mbh}{R},
  \end{equation}
  where $\Mbh$ is the mass of the SMBH.
  We define a mass ratio of SMBH, $\fbh$, which is evaluated by the mass of SMBH ($\Mbh$) divided by the total mass of the bulge 
  and the SMBH within $R$, i.e.
  \begin{equation}
    \fbh(R)=\frac{\Mbh}{M_{\rm g} (R) + \Mbh}.
  \end{equation}
  Thus, we can obtain $\Mbh$ from Equation~\ref{eq:mgr} and $\fbh(\Rg)$ as:
  \begin{equation}
    \Mbh = \frac{4\pi \rhogz \Rg^3 \fbh(\Rg) }{1-\fbh(\Rg)} 
    \begin{cases}
      \frac{1}{ 3-\alpha} & \text{for}~\alpha \ne 3\\
      \log{\Rg} & \text{for}~\alpha =3
    \end{cases}
  \end{equation}

  For two-component (galaxy+SMBH) potentials, we can also find a similar boundary between tidal disruption and compression like Equation~\ref{eq:rtde}.
  Based on the disruption criterion \citep[Equation 36 in ][]{Ivanov2020} for star clusters with homogeneous density profile, tidal compression occurs with
  \begin{equation}
    \gamma/\Omega > (\Omega/\omega_0)^{-1/4}
  \end{equation}
  where 
  \begin{equation}
    \begin{aligned}
      \Omega^2 & = \frac{1}{R} \frac{\partial}{\partial R} \Psi \\ 
      \gamma^2 & =  4 \Omega^2 - \frac{2\Omega}{R} \frac{\partial}{\partial R} \left ( R^2 \Omega \right) \\
      \omega_0^2 & = \frac{4\pi}{3} G \rhogz .
    \end{aligned}
  \end{equation}

  \begin{figure}[ht!]
    \plotone{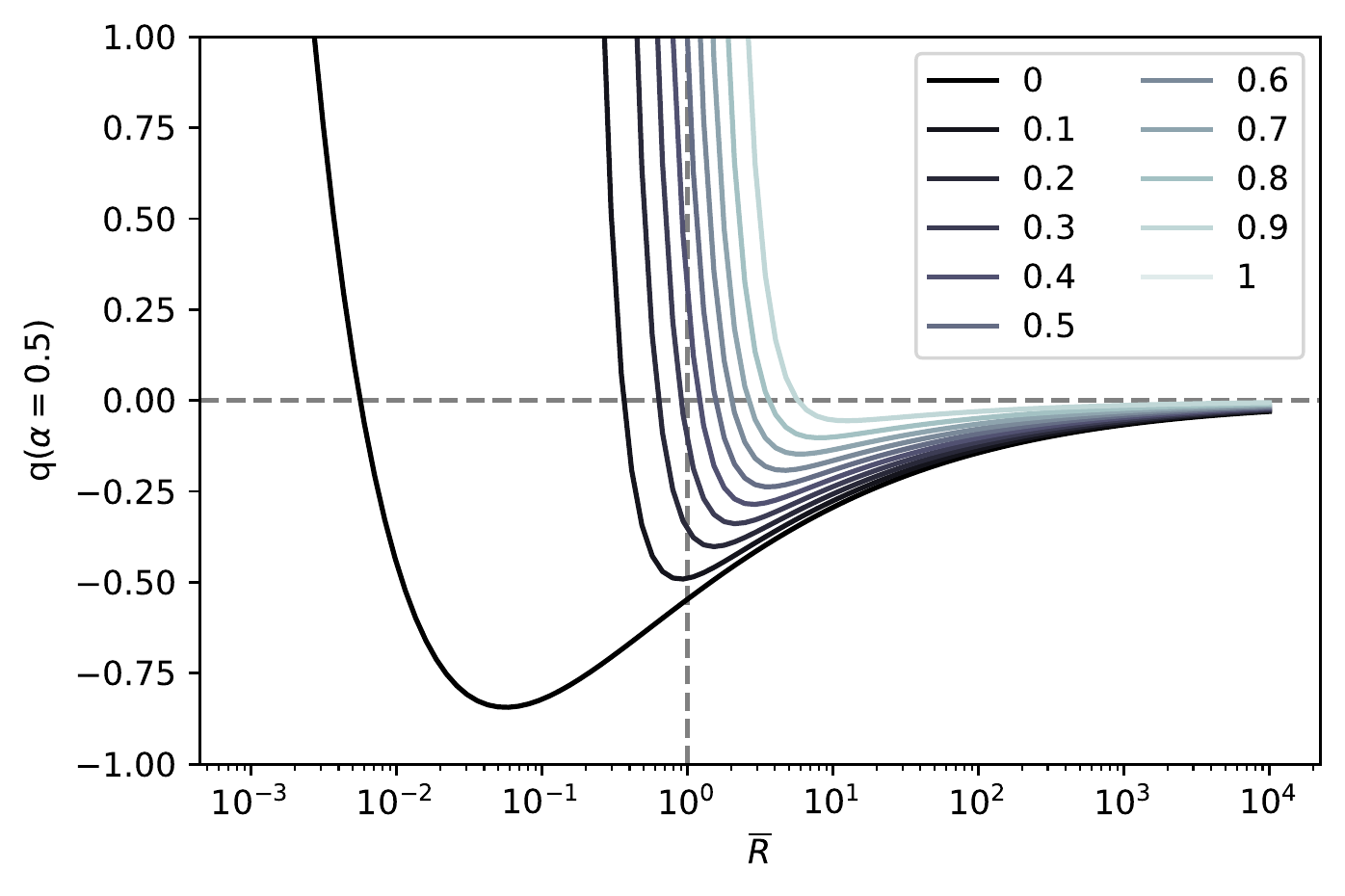}
    \caption{The $q-\Rs$ relation for $\alpha=0.5 $. Colors indicate $\fbh$ and the values are shown in the legend. 
      The overlap points between the $q-\Rs$ curves and the horizontal dashed line ($q=0$) indicates the $\Rs$ boundary of tidal disruption 
      ($q > 0$) and compression ($q<0$). 
      The vertical line ($\Rs=1$) indicates the position of star cluster in the two-component potential.
      If $q\le 0$ at $\Rs=1$, the cluster does not suffer tidal disruption.
      \label{fig:psbhrs}}
  \end{figure}

  For presentation convenience, we define
  \begin{equation}
    \Rs \equiv \frac{R}{\Rg},
  \end{equation}
  where $\Rs$ represents the dimensionless distance to the galactic center normalized in the unit of $\Rg$.
  For a power-law galactic potential, 
  \begin{equation}
    \begin{aligned}
      \frac{\Omegag^2}{\omega_0^2} & = 
                                     \begin{cases}
                                       \frac{3 \Rs^{-\alpha}}{3 - \alpha} & \text{for}~\alpha\ne 3 \\
                                       \frac{3\ln{\Rs}}{\Rs^3} &  \text{for}~\alpha =3
                                     \end{cases} \\
      \gammag^2 &= \alpha \Omegag^2
    \end{aligned}
    \label{eq:omegamma}
  \end{equation}
  where $\Omegag=\Omega(Rg)$, and $\gammag=\gamma(\Rg)$.
  For the point-mass (SMBH) potential at the same $R$,
  \begin{equation}
    \begin{aligned}
      \Omegabh^2 &= \frac{ G \Mbh}{R^3} \\
      \gammabh^2 &= 3 \Omegabh^2
    \end{aligned}
  \end{equation}
  With the two components, the criterion of Equation~\ref{eq:rtde} can be written as
  \begin{equation}
    \frac{\fbh(\Rg) \gammabh^2 + \left (1-\fbh(\Rg) \right) \gammag^2}{\left [ \fbh(\Rg) \Omegabh^2 + \left (1-\fbh(\Rg) \right ) \Omegag^2 \right ]^{3/4}} > \omega_0^{-1/2}.
    \label{eq:fbh}
  \end{equation}
  After some algebra, this condition for tidal disruption (with $\alpha<\amax$) can be rewritten as 
  \begin{equation}
    \begin{aligned}
      q &= \frac{3 \overline{R}^{3 - \alpha} \alpha \left(\fbh(\Rg) - 1\right)^{2} + 9 ^{2}}{\overline{R}^{3} \left(\fbh(\Rg) - 1\right) \left(\alpha - 3\right)} \\
        &- \left(\frac{3 \overline{R}^{3 - \alpha} \left(\fbh(\Rg) - 1\right)^{2} + 3 \fbh(\Rg)^{2}}{\overline{R}^{3} \left(\fbh(\Rg) - 1\right) \left(\alpha - 3\right)}\right)^{\frac{3}{4}} > 0.
    \end{aligned}
    \label{eq:qcritical}
  \end{equation}
  The value of $R$ (or $\Rs$)
  at $q=0$ is equivalent to $\Rd$ in Equation (\ref{eq:rtde}) which again differs from
  the conventional tidal disruption radius of $R$ inferred from Equation (\ref{eq:rt}) with $\rt=r_{\rm cut}$.
  Hereafter, {\it $R <\Rd$ refers to the sufficient criterion for tidal disruption}.

\begin{figure}[ht!]
    \plotone{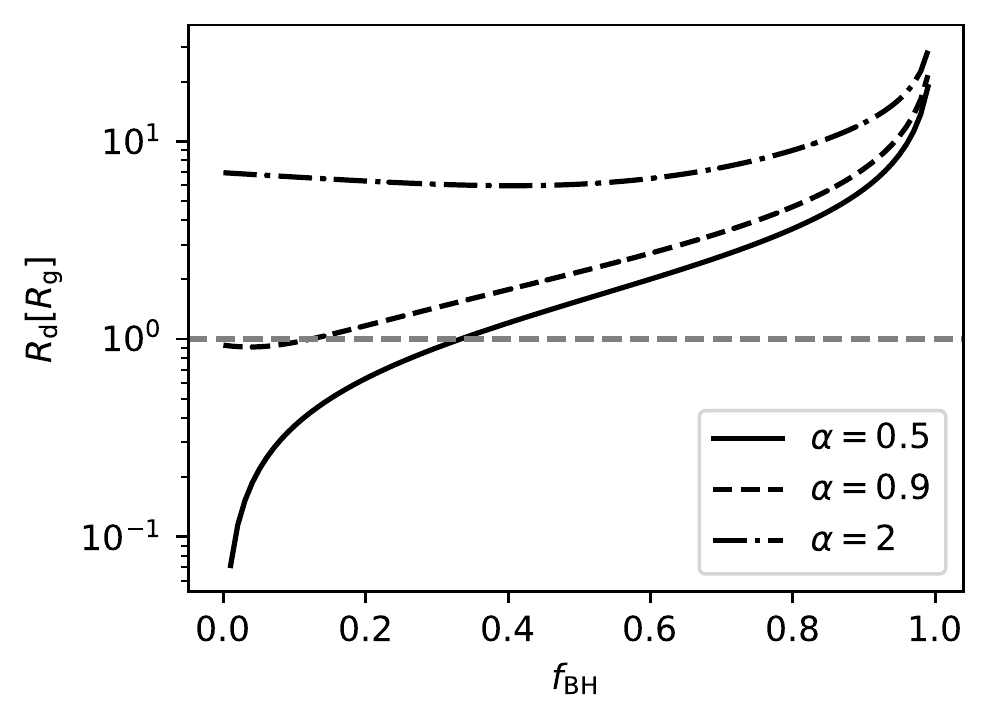}
    \caption{The $f_{\rm BH} (\equiv \fbh)-\Rd$ relation of the two-component potential. Three values of $\alpha$ are shown. When 
      $f_{\rm BH}$ is larger than the overlap point between the dashed line ($\Rd=\Rg$) and the curves, the cluster suffer tidal disruption. 
      For $\alpha=2$, the cluster always suffers tidal disruption. \label{fig:rdfbh}}
  \end{figure}

  We calculate $q-\Rs$ relation with $\alpha=0.5 \ (< \amax)$ for different $\fbh(\Rg)$, as shown in Figure~\ref{fig:psbhrs}.
  The location of $\Rd$ indicates the outer boundary of the tidal disruption domain.
  When $\fbh(\Rg) < 0.3$, $\Rd < \Rg$, and thus, the cluster does not suffer tidal disruption. 
  Figure~\ref{fig:rdfbh} shows the relation of $\Rd$ and $\fbh(\Rg)$ for three values of $\alpha$ (0.5, 0.9, and 2).
  For $\alpha=0.5$, the corresponding $\fbh(\Rg) = 1/3$ for $\Rd=\Rg$.
  With relative low SMBH-galaxy mass ratios, there is a range of galacto-centric distance from the SMBH smaller than that leads to cluster's Roche-lobe overflow
  (the conventional necessary condition for tidal disruption), where the tidal perturbation on the cluster is compressive.
  For $\alpha=0.9$, which is close $\amax$ in the case of the single power-law potential, the minimum $\Rd \simeq \Rg$ and
  $\fbh(\Rg)\approx 0.1$ when $\Rd=\Rg$.
  For $\alpha=2$, $\Rd>\Rg$ for all region of $\fbh(\Rg)$ such that star clusters always suffer tidal disruption
  in this galactic+SMBH potential  
  even in the limit of small SMBH-galaxy mass ratio.  The sufficient criterion for tidal disruption would be 
  satisfied when the necessary condition is met with $\Rd>\Rg$.

  To validate this prediction, we perform 4 $N$-body simulations of W2 star clusters under the two-component 
  (galaxy+SMBH) potential, where $\alpha=0.5$ and $\fbh(\Rg) = 0.01$, $0.1$, $1/3$, and $0.5$ respectively 
  (\S\ref{sec:psbh}). 
  The corresponding $\Rd/\Rg \approx 0.0711, 0.364, 1.00$ and $1.56$ for these values of $\fbh(\Rg)$, respectively.
  Similar to the discussion in \S\ref{sec:onecomponent}, the small $\Rd$ value for small $\fbh(\Rg)$ 
  indicates the tidal compression can occur well inside the conventional galacto-centric tidal disruption distance.
  These clusters are again placed at $\Rg=100~\rt$ with a circular orbit. 
  We also perform another group of simulations with the same two-component potential, 
  but placing the clusters on a circular orbit with different $\Rg$'s.
  This group of models represents a set of more general condition, where the values of 
  $\Rg$ are independently specified, hence the ratio $\rt/\rcut$ differs from unity.
  Since $\fbh(\Rg)$ increases when $\Rg$ decreases with the same potential, 
  these models also have different $\fbh(\Rg)$.
  The differences from the previous model set are that the clusters suffer a stronger tidal force and their orbital 
  period is shorter when $\Rg$ is smaller. The cut-off radius $r_{\rm cut}$ of this set of cluster models need not be the tidal radius $\rt$ (Eq. \ref{eq:rt}). 
  In the limit $\Rg<100~\rt$, a fraction of the cluster stars are initially placed inside $r_{\rm cut}$ but outside 
  $\rt$. With this general prescription, we investigate how the tidal effect determines the dynamical evolution
  of these stars. 
  The initial value of $\Rg$ and corresponding $\fbh(\Rg)$ are shown in Table~\ref{tab:varyR}.
  The results of the two model sets are shown in Section~\ref{sec:psbh}.

  \begin{table}[]
    \centering
    \begin{tabular}{c|ccccccc}
      \hline
      $\Rg[\rt]$ &  5 & 10 & 20 & 40 & 60 & 80 & 100\\
      $\fbh(\Rg)$ & 0.948 & 0.762 & 0.361 & 0.091 & 0.035 & 0.017 & 0.010 \\
      \hline
    \end{tabular}
    \caption{$\Rg$ and the corresponding $\fbh(\Rg)$ for the star cluster models under a two-component potential ($\alpha=0.05$).}
    \label{tab:varyR}
  \end{table}

  \subsubsection{Time-dependent background potential for clusters with slowly decaying orbit}
  \label{sec:dfriction}

  Star clusters undergo orbital decay due to dynamical friction and endure increasingly strong external potential.
  In order to represent this effect, we smoothly increase the scaled mass which contributes to the galactic potential
  in the co-moving frame centered on the cluster. Note that this generic scaling
  prescription is equivalent to the evolving external tidal field imposed on an in-spiraling stellar cluster 
  in a static galactic potential.  It does not correspond an actual (physical) mass gain for the galaxy and the SMBH.
  For a reference, we also integrate the orbit of a point-mass satellite in the single component potential with the Chandrasekhar’s dynamical friction, the results are presented and discussed in Appendix.
  For the time-dependent two-component potential, the masses of the SMBH and 
  the galactic bulge should increase with the same rate to correctly represent the static potential.
  But this scaling prescription does not mean that $\fbh(R)$ is a constant of radius. 
  As star clusters approach to the center, $R$ decreases while $\fbh(R)$ increases.

  We consider an evolved power-law potential  (with $\alpha=0.5$) and an evolved two-component potential (with 
  $\alpha = 0.5$ and $\fbh(\Rg)=0.01$)  where $\rhogz$ increases exponentially as
  \begin{equation}
    \rhogz(t) = \rhogz(0) e^{\rrho t}.
    \label{eq:rrho}
  \end{equation}
  By changing $\rrho$, we can investigate clusters with different in-spiraling time scales (\S\ref{sec:varyt}).
  When $\rrho$ is small, $\trh$ is less than the in-spiral time and star cluster follows a tightly wrapped spiral
  pathway.  We investigate three values of $\rrho$:  0.033, 0.065 and 0.13 $\trhz^{-1}$, 
  for both power-law and two-component potentials.
  Meanwhile, when massive GC's (with the same internal density) sink into the galactic center
  via dynamical friction, their inspiral time scale can be much shorter than its $\trh$.
  Thus, we include three additional models with the power-law potential and $\rrho$ of 4.2, 8.3, 17 $\trhz^{-1}$.
  The results of simulations are shown in \S~\ref{sec:varyt} (Fig. \ref{fig:kingpselagr2}).

  Our $N$-body models assume initial number of stars, $N=1000$.
  To validate that the tidal compression feature is independent of $N$, we also include 
  a group of simulations with $N=10000$ and the same total mass (Fig. \ref{fig:kingpselagr}).
  The clusters move in power-law potentials with $\alpha=0.5$ and $\rrho = 0, 0.17$ and $21 \trhz^{-1}$,  where $0$ represents a static potential. 
  The initial position is the same where $\Rg=100\rtz$.

  \subsubsection{Time-dependent background potential for clusters with highly eccentric orbits}
  \label{sec:eccinit}

  We investigate two sets of clusters moving on elliptical orbits around 
  some static (both power-law and two-component) tidal potentials.
  We adopt a modest eccentricity in the first set by setting the initial tangential 
  velocity of the cluster to be half of that in a circular orbit. We consider a 
  nearly parabolic set by setting a small initial tangential velocity 
  (about 8 percent of the circular one).   
  Table~\ref{tab:init} summarizes the model parameter of all simulations in this study. 
  The results of these simulations are shown in \S~\ref{sec:ecc}.

  \begin{table*}[]
    \centering
    \begin{tabular}{l|l|l|l}
      \hline
      Profile & Potential & Varying parameters & Number of models\\
      \hline
      Homo.  & power-law & $\alpha= 0.5, 1, 2$, point-mass, NoTide & 5\\
      \hline
      King  & power-law & $\alpha= 0.5, 1, 2$, point-mass, NoTide & 15\\
              &           & $W_0 = 2, 6, 8$ for each galactic potential \\
      \hline
      King  & power-law & $\alpha= 0.5, 1, 2$, point-mass, NoTide & 30 \\
      $W_0=2$ &           & 6 groups of different random seeds & \\
      \cline{2-4}
              & two-component & $\fbh(\Rg) = 0.01, 0.1, 1/3, 0.5$ & 4\\
      \cline{3-4}
              & $\alpha=0.5 + \fbh=0.01$     & $\Rg[\rt] = 5, 10, 20, 40, 60, 80, 100$ & 7\\
      \cline{2-4}
              & secular decay, $N=1000$& $\rrho = 0.033, 0.065, 0.13 [\trhz]$ & 6 \\   
              &               & $\alpha=0.5$, $\alpha=0.5+\fbh=0.01$  & \\
      \cline{3-4}
              &               & $\rrho= 4.2, 8.3, 17~[\trhz]$ & 3\\
      \cline{2-4}
              & secular decay, $N=10000$ &  $\rrho = 0.033, 0.065, 0.13~[\trhz]$ for $\alpha=0.5$, NoTide & 4 \\
      \cline{2-4}
              & eccentric orbit     & eccentric, nearly-radial & 4\\
              &               & $\alpha=0.5$, $\alpha=0.5+\fbh=0.01$ & \\ 
      \hline
    \end{tabular}
    \caption{Initial conditions of all simulations. "Homo." and "King" represent the homogeneous and \cite{michie1963}-\cite{king1966} profiles, respectively. $\alpha=0.5$ and $\alpha=0.5+\fbh=0.01$ represent the power-law potential with $\alpha=0.5$ and the two-component potential with $\alpha=0.5$ and $\fbh(\Rg)=0.01$, respectively.}
    \label{tab:init}
  \end{table*}

  \section{Clusters on nearly circular orbit around the galaxy}
  \label{sec:circular}

  \subsection{Static one-component galactic potential}
  \subsubsection{Homogeneous sphere}
  \label{sec:homo}

\begin{figure*}[ht!]
    \plotone{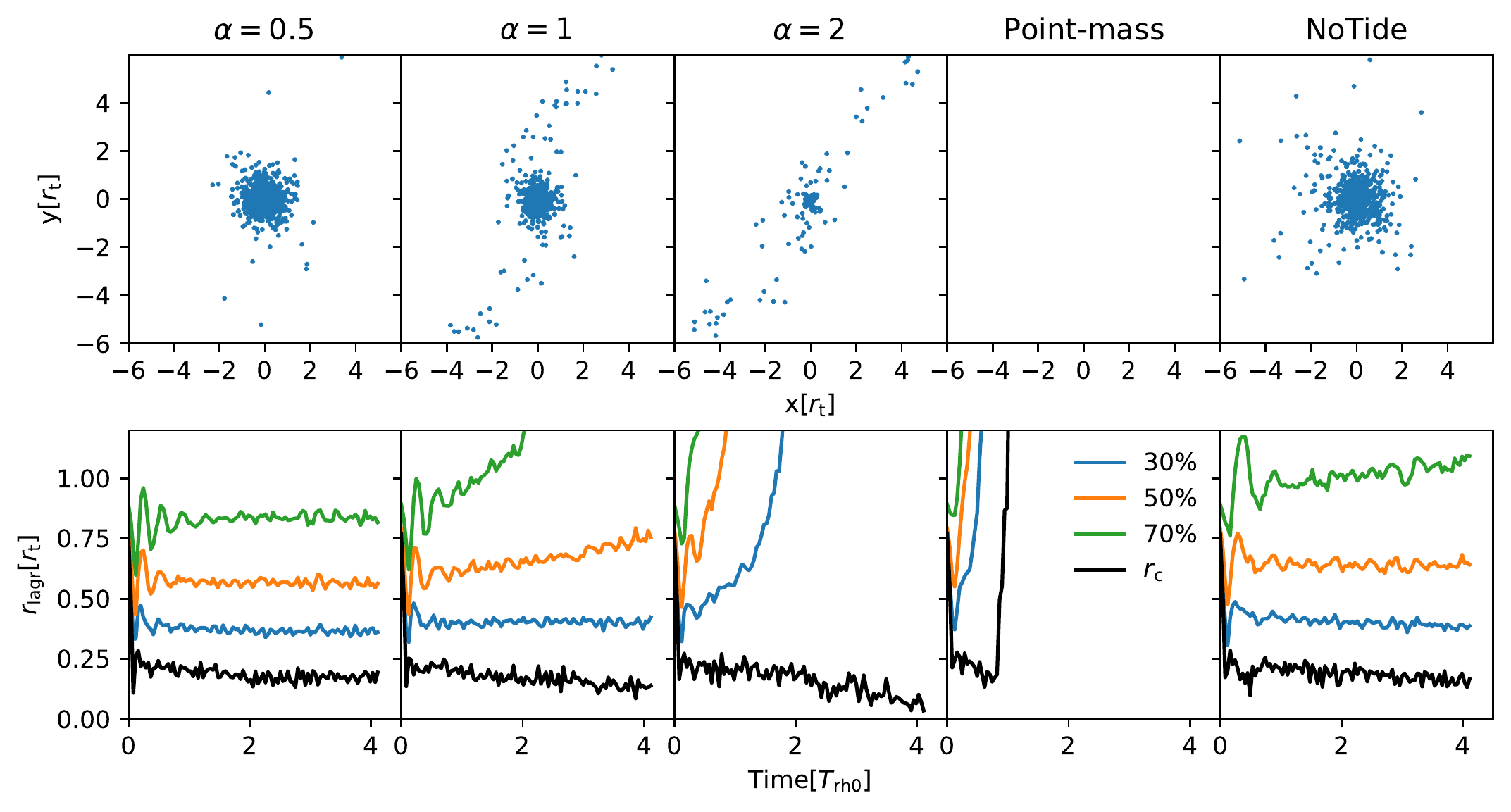}
    \caption{The morphology at about $4~\trhz$ (upper panel) and the evolution of Lagrangian radii ($\rl$) and core radii ($\rc$) of star clusters with the spherical symmetric homogeneous mass-density distribution (lower panel). The columns show different galactic potentials. The first three represent power-law spherically symmetric potentials with different $\alpha$ (see Equation~\ref{eq:ps}). The last two show the point-mass potential and the reference without tidal field (NoTide). Colors in the lower panels indicates the mass fractions of $\rl$, except that the red color represents the core radius $\rh$. \label{fig:homo}}
  \end{figure*}

  Figure~\ref{fig:homo} shows the morphology of star clusters with the homogeneous sphere 
  (\S\ref{sec:homogenous}) in a power-law potential (\S\ref{sec:onecomponent}) at $4~\trhz$ and 
  the evolution of different Lagrangian radii ($\rl$) and the core radius ($\rc$). We separate the 
  tidal response from the clusters' internal two-body-relaxation effects using comparisons 
  between clusters embedded in the external tidal field and the appropriate NoTide calibration 
  model.  For these isolated star clusters (last panel), the system's core contracts within 
  $0.1~\trhz$ or $2~\tcrz$ ($\trhz \approx 18.3~\tcrz$, where $\tcrz$ is the initial crossing 
  time in Eq. \ref{eq:crossingtime}), then expands and virializes to its original structure 
  after about $4~\tcrz$. During the subsequent quasi-stable relaxation, the inner 
  radii ($< \rh \equiv \rl(50\%)$, the half-mass radius) contract while the outer radii ($> \rh$) expand.  This 
  evolution is a direct consequence of two-body relaxation\citep{Spitzer1987, heggie2003} and we use this model for
  calibrator to accentuate the tidal influence.  The effects of tidal compression and disruption
  respectively leads to the deceleration and acceleration in the expansion of the outer $\rl>r_{\rm h}$.

  In the proximity of a point-mass potential, clusters with $r_{\rm cut} \gtrsim \rt$ quickly suffers 
  total tidal disruption within one $\trhz$. For single power-law potentials with various values 
  of $\alpha$ (Eq. \ref{eq:ps}), the clusters become less vulnerable to tidal disruption in the limit of 
  small $\alpha$. In the case of $\alpha=1$, which is close to $\amax$, the radii $\rl$ within the half mass radius $\rh$ follow similar 
  evolutionary paths as those in the NoTide calibration model without the galactic potential.
  Nevertheless, the outer radii for this model show modest amount of tidal removal.
  In contrast, the model with $\alpha=0.5$ shows the characteristics of tidal compression, where outer radii 
  do not expand. For the $\alpha=2$ model and the point-mass potential, both the sufficient 
  and necessary conditions for tidal disruption are satisfied and the cluster's $\rl$ expands faster 
  than their equivalent radii for the NoTide calibration model.
  This $\alpha$-dependence is consistent with the analytic calculation carried out by \cite{Ivanov2020}.

  \subsubsection{Clusters with a Michie-King initial density distribution}
  \label{sec:kingw0}

  \begin{figure*}[ht!]
    \plotone{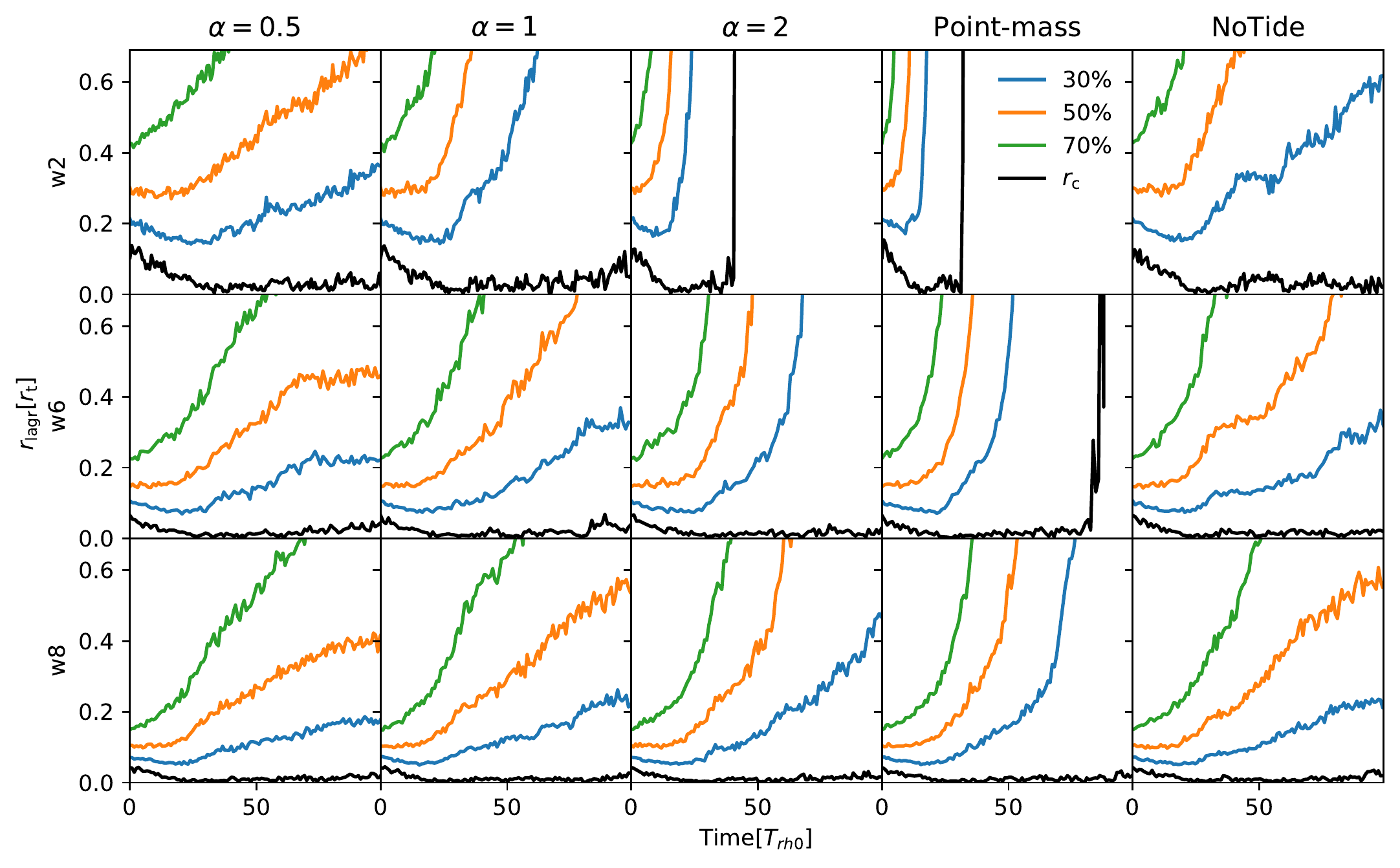}
    \caption{The evolution of Lagrangian radii of star clusters with different $W_0$ of King model. The galactic potential set is the same as that in Figure~\ref{fig:homo}. \label{fig:kingw0ps}}
  \end{figure*}

  Figure~\ref{fig:kingw0ps} show the evolution of $\rl$ and $\rc$ for models with Michie-King profiles 
  (\S\ref{sec:michie}) in the clusters and power-law potentials (\S\ref{sec:onecomponent}) for the host
  galaxy.  All models are evolved for more than $80~\trhz$.  In the absence of galactic potential, 
  the NoTide calibration model shows core collapse after the first $20~ \trh$.  During the long-term 
  evolution in the post-core-collapse phase, all $\rl$ expands because binary systems form in the 
  core and stars are ejected from the core through binary-single-star close encounters.

  For the $\alpha=2$ as well as the point-mass galactic potential, the clusters with small $W_0$ are rapidly 
  disrupted.  Similar to the case in Figure~\ref{fig:homo}, the evolution for $\alpha=1$ power-law potential 
  is close to that without galactic potential. Although the half-mass radius $r_{\rm h}( \equiv \rl (50\%))$ 
  in the W2, W6, and W8 clusters with $\alpha=0.5$ galactic potential increases with time, its expansion
  rate is slower than that in the NoTide calibration model.  This evolutionary pattern is the characteristics 
  of tidal compression. Comparison between Figures \ref{fig:homo} and \ref{fig:kingw0ps} indicate that
  the internal density profile of star clusters does not significantly affect the criterion for tidal 
  disruption and compression (Equation~\ref{eq:rtde}).  Since the structure evolution of the W2 model 
  is most sensitive to the galactic potential, we apply this density profile as a default model for the 
  star clusters in the following analysis with different galactic potentials.

  \begin{figure}[ht!]
    \plotone{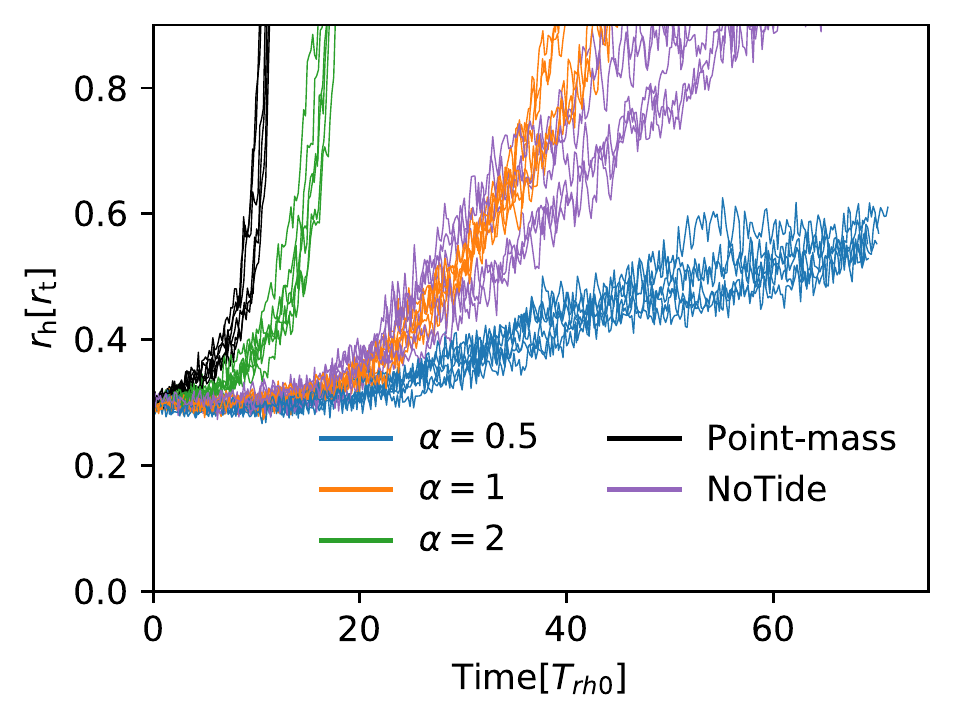}
    \caption{The evolution of $\rh$ of star clusters with $W_0=2$, different galactic potentials and 6 different sets of 
      random seeds for initial conditions. Each curve represents one simulation. Curves with the same color have the same 
      potential. \label{fig:randcheck}}
  \end{figure}

  Models with the same density profile (i.e. identical initial stellar positions and velocities)
  in Figure \ref{fig:homo} and \ref{fig:kingw0ps} are embedded in
  different galactic potentials.
  Keeping the same initial condition can reduce the stochastic scatter due to the low number of stars.
  To confirm that the tidal compression is robust, we perform a set of tested models 
  with 6 different random seeds, and compare how $\rh$ evolves under different galactic potential.
  The results in Figure \ref{fig:randcheck} show that the stochastic scatter due to the random seeds is much 
  smaller than the difference caused by tidal effect imposed by various prescriptions for the galactic potentials. 

  \subsection{A two-component model for the galaxy-SMBH composite potential}
  \label{sec:psbh}

\begin{figure*}[ht!]
    \plotone{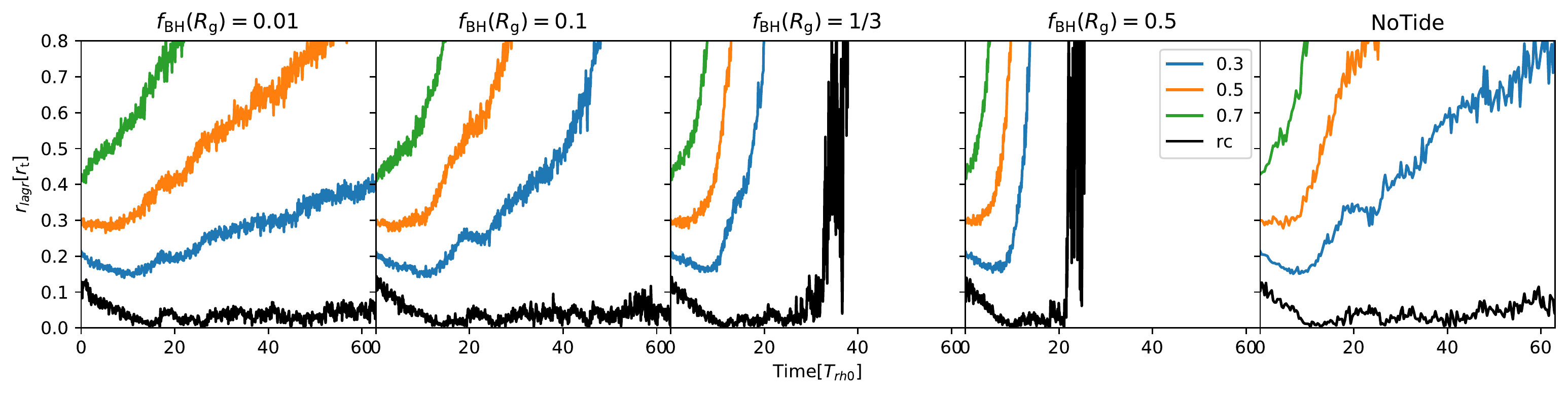}
    \caption{The evolution of Lagrangian radii of star clusters in the two-component potential with $\alpha=0.5$. Columns show different mass
      ratios between SMBH and galactic background. The model without galactic potential is shown in the last column. \label{fig:kingpsbh}}
  \end{figure*}

  Figure~\ref{fig:kingpsbh} shows the evolution of $\rl$ and $\rc$ for the two-component models of 
  $\alpha=0.5$ with four values of $\fbh$ (\S\ref{sec:twocomponents}).  Based on their comparison with
  the NoTide calibration model, the clusters surrounded by the $\fbh=0.01$ and $0.1$ potential 
  are tidally compressed.  For relative high SMBH-galaxy mass ratios (with $\fbh=1/3$ 
  and $0.5$ at $\Rd=\Rg$), the sufficient criterion for tidal disruption is satisfied when the
  conventional, Roche-lobe-filling, necessary condition is met and the cluster's $\rl$ expand faster than their 
  equivalent radii for the NoTide calibration model.
  These results are consistent with the value of $\fbh$ for 
  the $\Rd=\Rg$ transition inferred from Figure~\ref{fig:rdfbh} and Equation (\ref{eq:qcritical}). 
  The evolution of $\rl$ for both homogeneous and Michie-King profiles show that the outermost $\rl$ is most 
  sensitive to the tidal effect.  To simplify the comparison, for the following analysis of tidal effect, 
  we only show the evolution of $\rl$ with the mass fraction of 90 percent.

\begin{figure*}[ht!]
    \plotone{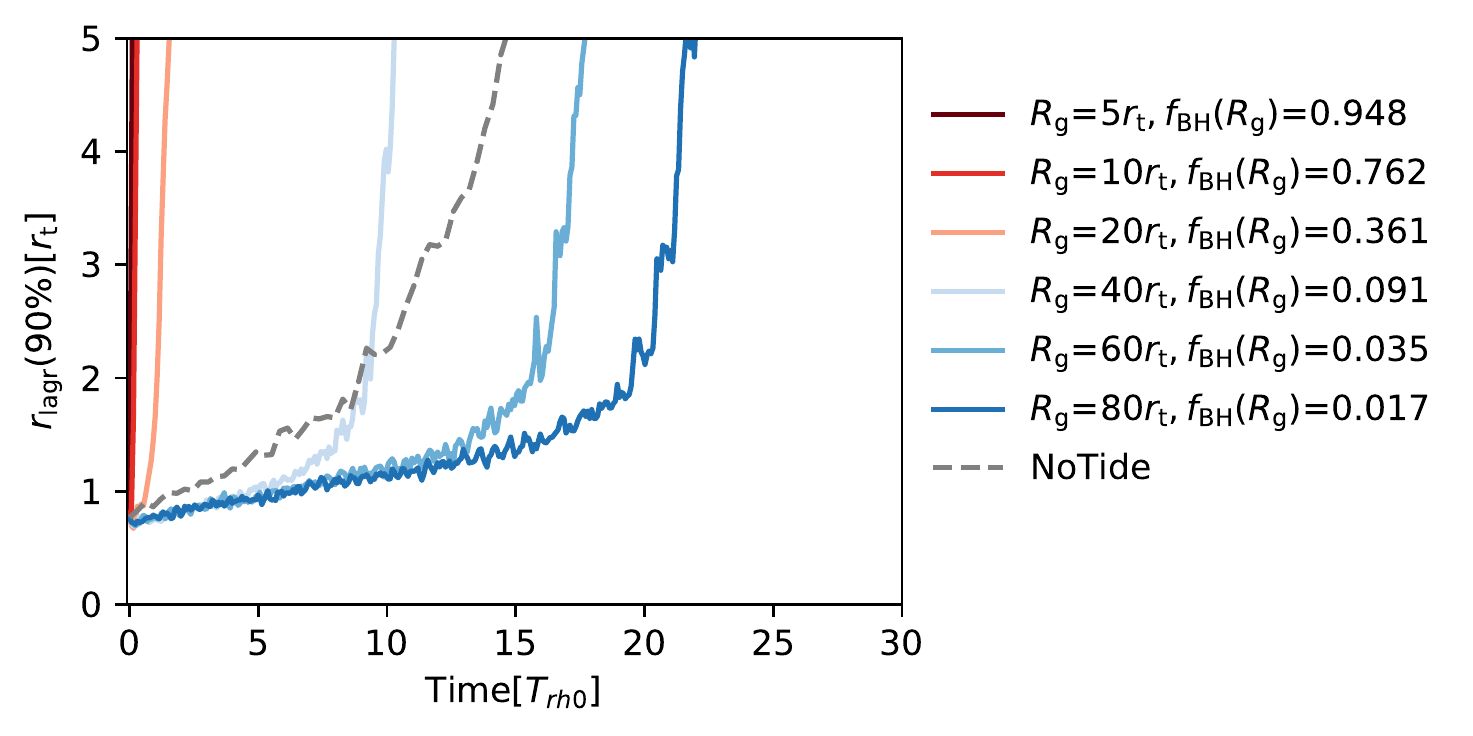}
    \caption{The evolution of $\rl(90\%)$ for star clusters in the two-component potential with different $\Rg$ (see Table~\ref{tab:varyR}). The NoTide model is shown as dashed curve for a reference. 
      \label{fig:varyR}
    }
  \end{figure*}

  When $\Rg=100~\rcut$, a special condition occurs 
  in which star cluster's $\rt$ overlaps $\rcut$ of the King profile. We consider more general cases with this special condition as a reference point. 
  In models with $\Rg<100~\rcut$, a fraction of the 
  cluster stars are initially located outside 
  $\rt$ (Table~\ref{tab:varyR}).  As $\Rg$ changes, the corresponding $\fbh$ also varies. In the range 
  $\Rg\ge 40~\rt$ ($\fbh < 1/3$), $\rl(90\%)$ increases slower than that of the NoTide calibration
  model (Fig.~\ref{fig:varyR}).  This difference suggests a tidal compression process 
  which is also consistent with critical value of $\fbh$ for the $\Rd=\Rg$ transition (Fig.~\ref{fig:rdfbh}). 
  After about $8~\trhz$, $\Rg \le \Rd \simeq 40~\rt$, $\rl(90\%)$ of increases faster than that of the 
  NoTide calibration model, which suggests a tidal disruption feature.  Similar values of $\Rd$ are
  obtained for models with larger initial $\Rg$, albeit the transition take place at later times.
  These results suggest that stars outside $\rt$ suffer both tidal compression and disruption effects at different stages. 

  \begin{figure*}[ht!]
    \plotone{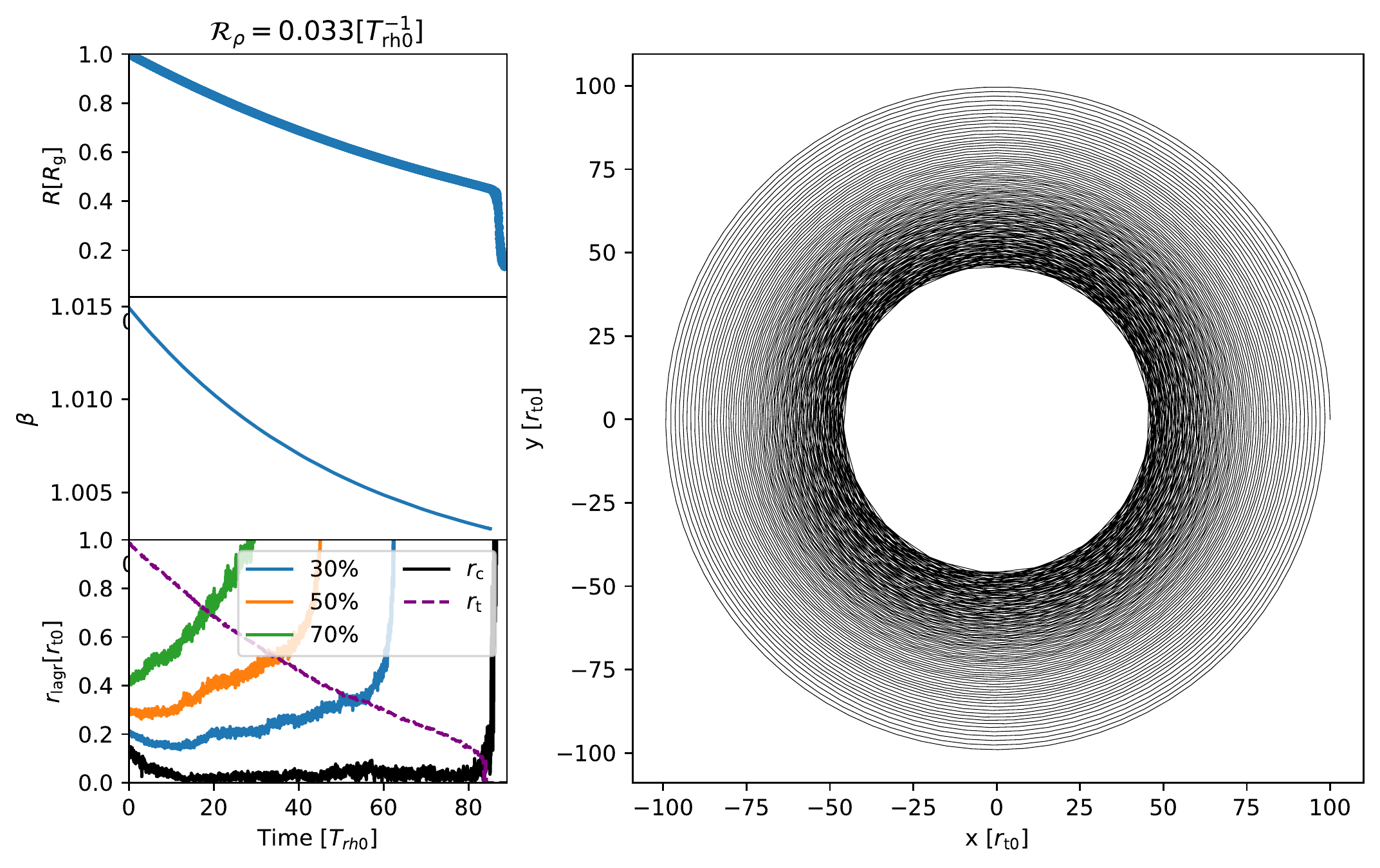}
    \caption{In the left panel from the top to the bottom: the distance ($R$) to the galactic center, $\beta$ from Equation~\ref{eq:rrhot} with the $P$ calculated at $\rt$, and the evolution of $\rl$  and $\rc$. The star cluster evolve in an time-dependent power-law potential with $\alpha=0.5$ and $\rrho\approx0.033~\trhz^{-1}$. In the right panel: the trajectory of the star cluster in the orbital plane of the galactic frame. The length unit $r_{\mathrm{t0}}$ is the initial $\rt$.  
      After the complete tidal disruption where $\rc \gg \rt$, no physical core exists, and a transition in the orbital trajectory appears near $R = 0.4~\Rg$.
      \label{fig:kingpse1}}
  \end{figure*}

  \section{Time-dependent potential due to orbital decay}
  \label{sec:evolve}

  In this section, we investigate clusters' evolution in some time-dependent potentials including those 
  due to their inward migration or due to their eccentric orbits.  For the former case, we artificially 
  increase the scaled masses of the galaxy and the SMBH as a generic approximation to the effect 
  of orbital decay due to dynamical friction (\S\ref{sec:dfriction}).  The physical mass and potential 
  distribution of the host galaxy that the model represents do not actually change over time. 
  To characterize the star clusters' initial structure, we choose the W2 density profile 
  since its response to the galactic tidal effect is more pronounced. 

  \subsection{Clusters with monotonic secular orbital decay}
  \label{sec:varyt}

  \begin{figure*}[ht!]
    \plotone{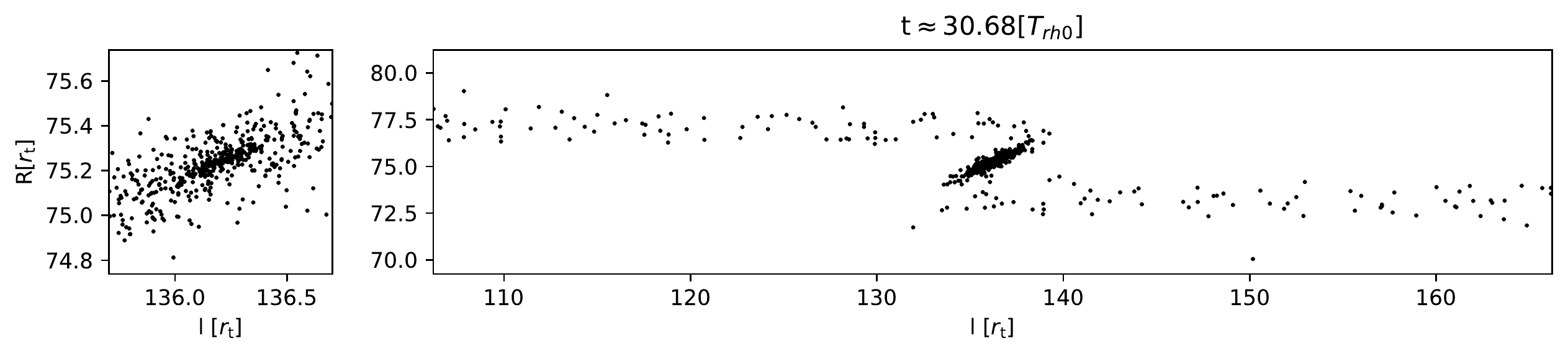}
    \caption{The projected morphology of the star cluster shown in Figure~\ref{fig:kingpse1} at about $30.68~\trhz$. The polar coordinate in the galactic orbital plane ($x$-$y$) is used, where $l=R \phi$ and $\phi$ is  angle. The left panel is the zoomed-in image. \label{fig:kingpse1orbit}}
  \end{figure*}

  We investigate how $\rrho$ affects the evolution of star clusters under the time-dependent potential described by 
  Equations \ref{eq:rrho} and \ref{eq:rrhot}.  To guide our interpretation, 
  we first analyze how the local density of the galactic background 
  varies as a function of $\rrho$.
  We define $\Delta \rhog$ to be the maximum variation in the circumscribed galactic mass density as a typical star revolves around 
  a cluster.  For a static power-law galactic potential, it can be approximated as:
  \begin{equation}
    \begin{aligned}
      \Delta \rhog & \simeq \rhogz \left [ \left ( \frac{\Rg}{R-\Delta R} \right)^\alpha - \left ( \frac{\Rg}{R+\Delta R} \right)^\alpha \right ] \\
                   & \approx 2 \alpha \rhogz  R_{g}^\alpha \Delta R  \frac{1}{R^{\alpha+1}}
    \end{aligned}
    \label{eq:drho}
  \end{equation}
  where $\Delta R$ is the maximum difference of $R$ for the star (traveling from the minimum of to the maximum of $R$), and 
  the approximation of $\Delta \rhog$ is derived by taking the first-order Taylor expansion of Equation \ref{eq:ps}.  If $R$ does 
  not change, i.e. $\Delta R$ changes sign throughout each orbit, the net amount of $\Delta \rhog$ would vanish.

  For a time-dependent potential resulting from the cluster's orbital decay, clusters travel 
  from $R - \Delta R$ to $R + \Delta R$ with a finite $\Delta R \propto R^{-\beta}$ where 
  $\beta=e^{\rrho P/2}$ during each orbital period $P$. The corresponding finite net residual
  \begin{equation}
    \begin{aligned}
      \Delta \rhog & = \rhogz \left [ \left ( \frac{\Rg}{R-\Delta R} \right)^\alpha - e^{\frac{\rrho P}{2}}  \left ( \frac{\Rg}{R+\Delta R} \right)^\alpha \right ] \\
                   & \approx \rhogz  R_{g}^\alpha \left [ \frac{1-\beta}{ R^\alpha} + \left (1 + \beta \right) 
                     \frac{\alpha \Delta R}{R^{\alpha+1}} \right ].
    \end{aligned}
    \label{eq:rrhot}
  \end{equation}
  In the limit $\rrho =0$ and $\beta = 1$, Equation(\ref{eq:rrhot}) reduces to Equation(\ref{eq:drho}) with no net 
  residual $\Delta \rhog$.  Provided $\beta$ is slightly larger than unity, stars in the clusters adjust to 
  adiabatic modifications in the strength of the host galaxy's tidal potential.

  \subsection{Tidal disruption along the course of orbital decay}
  Figure~\ref{fig:kingpse1} shows the evolution of $R$, $\beta$, $\rl$, $\rc$ and $\rt$ for the  model with a power-law potential in which $\alpha=0.5$, $\fbh(\Rg)=0.01$, and $\rrho \approx0.033~\trhz^{-1}$. 
  As the scaled mass of the galaxy increases, the star cluster spirals towards the  
  galactic center.  Note that the cluster's orbital trajectory obtained with idealized, 
  generic prescription qualitatively agrees with the numerical integration of the conventional dynamical friction
  formulae (Fig. \ref{fig:pmorbit}).
  As $R$ decreases, the cluster's corresponding $\rt$ shrinks.
  After the cluster's core collapse, the outer $\rl (\ge r_{\rm h})$ continues to 
  expand. During the entire course of evolution, $\beta$ is close to unity and decreases 
  monotonically (Eq. \ref{eq:rrhot}).  The system initially evolves in an adiabatic manner.
  As different $\rl$'s increase beyond $\rt$, their expansion rate accelerates.
  This pattern is an indication that stars outside $\rl$ have suffered tidal disruption,
  analogous to that of models with $\Rd=40~\rt$ in Figure~\ref{fig:varyR}.  Eventually, 
  when $\rc>\rt$, the cluster completely and the determination of the cluster center 
  becomes invalid. After core disruption, dynamical friction ceases to be effective
  \citep{fellhauer2007} and the evolution of $R$ can no longer represents the position 
  of the star cluster.

  \begin{figure}[ht!]
    \plotone{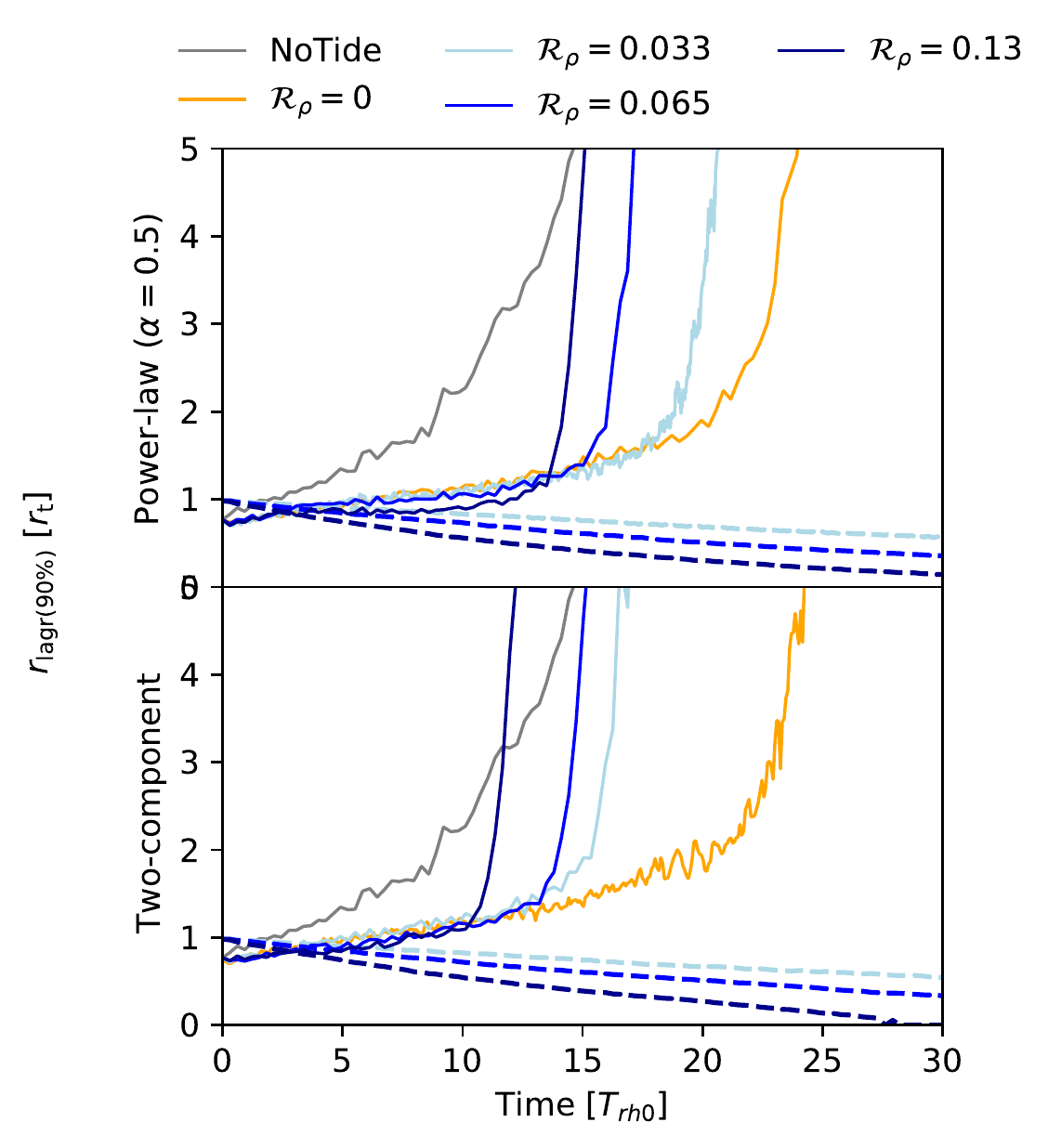}
    \caption{The evolution of  $\rl(90\%)$ (solid lines) and $\rt$ (dashed lines) for models with time-dependent power-law potential (upper panel) and two-component potential (lower panel). The reference models with static potentials and NoTide calibration model are shown as grey and orange colors, respectively. The value of $\rrho$ are in the unit if $\trhz^{-1}$. \label{fig:kingpselagr}}
  \end{figure}

  To understand how tidal disruption occurs in the limit $\rl>\rt$, we plot,
  in Figure~\ref{fig:kingpse1orbit}, the morphology of the cluster at about 
  $30.68~\trhz$ when the half-mass radius almost reaches $\rt$. Although
  the cluster is tidally compressed in the $R$ direction, it is also distorted along the tangential direction 
  with some phase lag and two tidal tails form. Thus, both radial compression and azimuthal disruption 
  determine the evolution of star clusters.
  This illustration also explains why the evolution of $\rl$ depends on the clusters' 
  orbits as shown in Figure~\ref{fig:varyR}.

  With a follow-up study, we compare models with different $\rrho$ and show the evolution of $\rl(90\%)$ and $\rt$ in 
  Figure~\ref{fig:kingpselagr}.  Here the three $\rrho$ values are small such that the cluster's in-spiral time 
  is much longer than $\trh$ and the cluster's orbits undergo tightly wrapped decay. 
  Both the time-dependent power-law potentials and the two-component potentials are investigated.
  For the two-component (galaxy+SMBH) potentials, both the scaled mass of the galaxy and the SMBH increase with the same $\rrho$ , 
  i.e, $\fbh (\Rg)$ does not change with time.
  In this case, as $R$ (or equivalently $\Rs = {R}/{\Rg}$) of the star cluster decreases, $\fbh (R)$ increases.
  The corresponding $R$ for $\fbh(R)=1/3$ is about $0.21~\Rg$.

  With a larger $\rrho (=0.13)$, $\rl(90\%)$ shows more pronounced characteristics 
  of tidal compression compared with the case of the NoTide calibration and the static ($\rrho =0$) power-law potential.  But the system also reaches the disruption phase earlier as $\rt$ shrinks faster than the model with smaller $\rrho$.
  In general, clusters dissolute slightly faster under the two-component (galaxy+SMBH) than the power-law galactic potential. 

  \begin{figure}[ht!]
    \plotone{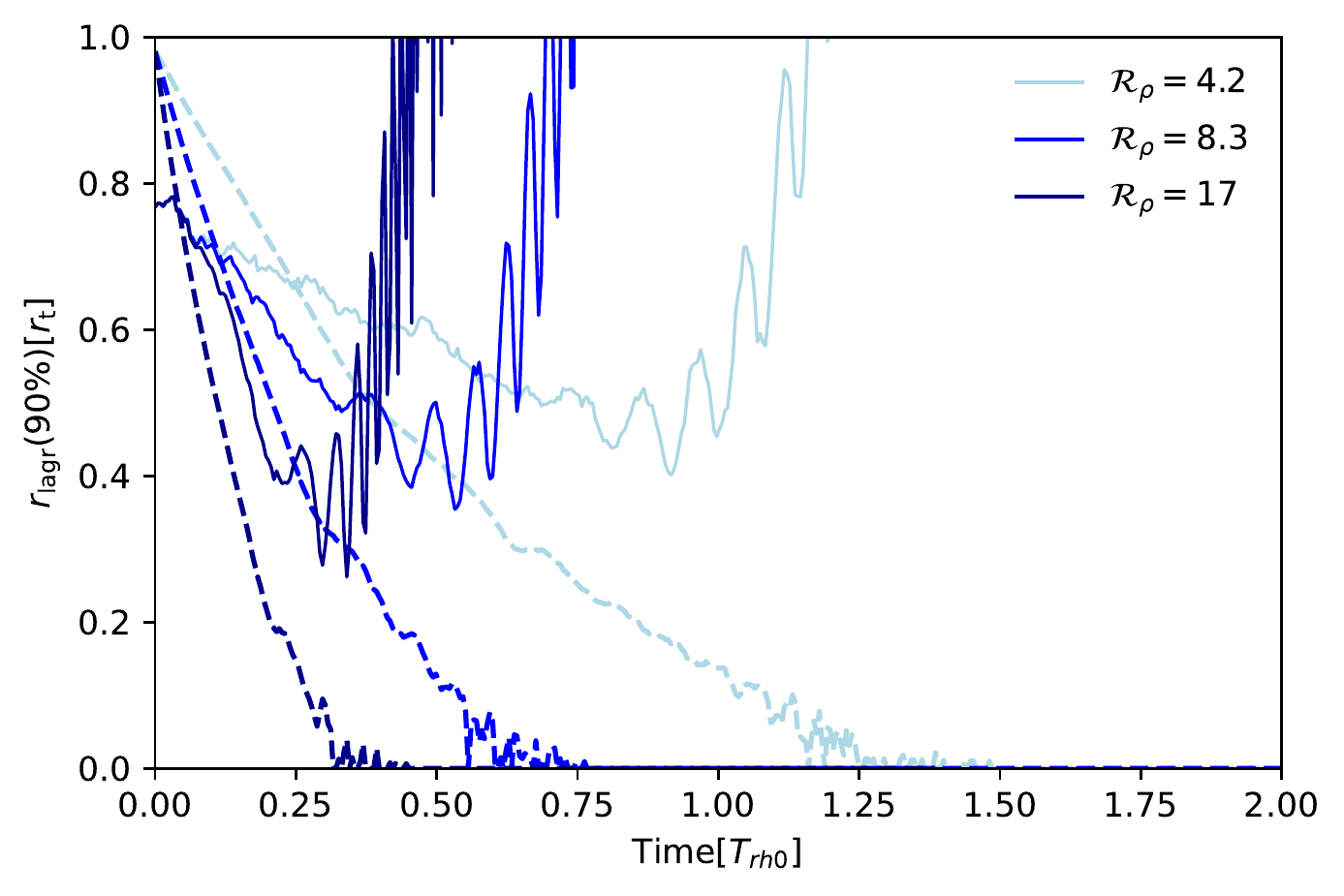}
    \caption{The solid and dashed lines represent the same quantities as in Figure~\ref{fig:kingpselagr}.
      Only models with power-law potential and large $\rrho$ are presented here.  \label{fig:kingpselagr2}}
  \end{figure}

  \begin{figure*}[ht!]
    \plotone{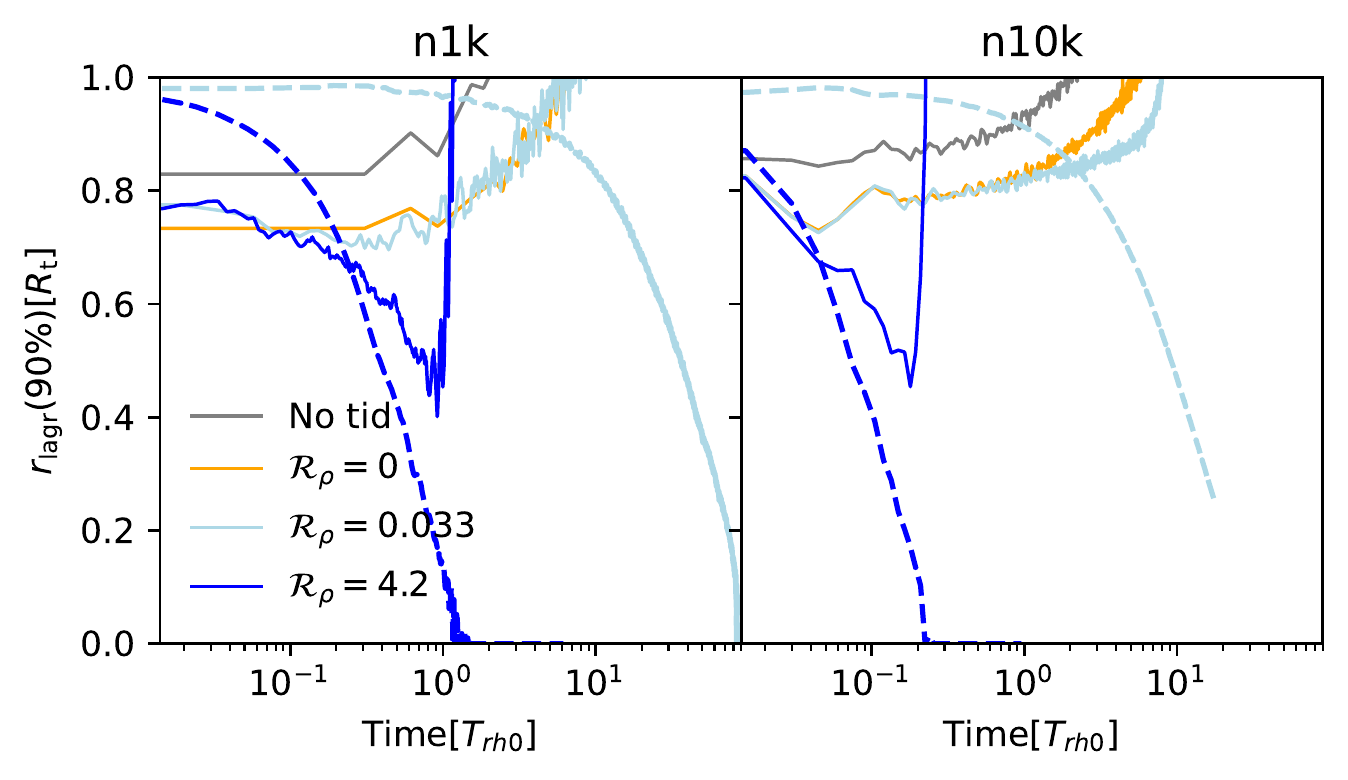}
    \caption{Comparing the models with different $N$. The plotting style is similar to Figure~\ref{fig:kingpselagr}. \label{fig:kingpse10k}}
  \end{figure*}

  Figure~\ref{fig:kingpselagr2} shows another set of three models with much larger $\rrho$, where $\trh$ is longer than the in-spiral time. 
  In these cases, $\rl(90\%)$ contracts slightly before any significant expansion.  This evolutionary pattern clearly indicates 
  the onset of tidal compression before disruption. Moreover, $\rl(90\%)$ continues to decrease for a while after $\rt$ has become 
  smaller than it.   This tendency suggests that tidal compression in the radial direction can cause some delays in 
  the tidal disruption of the cluster in the azimuthal direction and clusters in galaxies with sufficiently 
  small $\alpha$ can preserve their integrity after their orbits have decayed, with modest $f_\bullet (R)$, inside the conventional 
  Roche-lobe-filling tidal disruption radius. Nevertheless, they eventually disintegrate due to the tidal dispersal in the azimuthal direction.

  Figure~\ref{fig:kingpse10k} compares the models with different $N$.
  With 10 times larger $N$, the evolution of $\rl$ has nearly identical tendency 
  (including the NoTide calibration model), albeit the evolution timescale is 
  slight shorter for the $\rrho=4.2$ case.

  \subsection{Nearly-radial orbits}
  \label{sec:ecc}

\begin{figure*}[ht!]
    \plotone{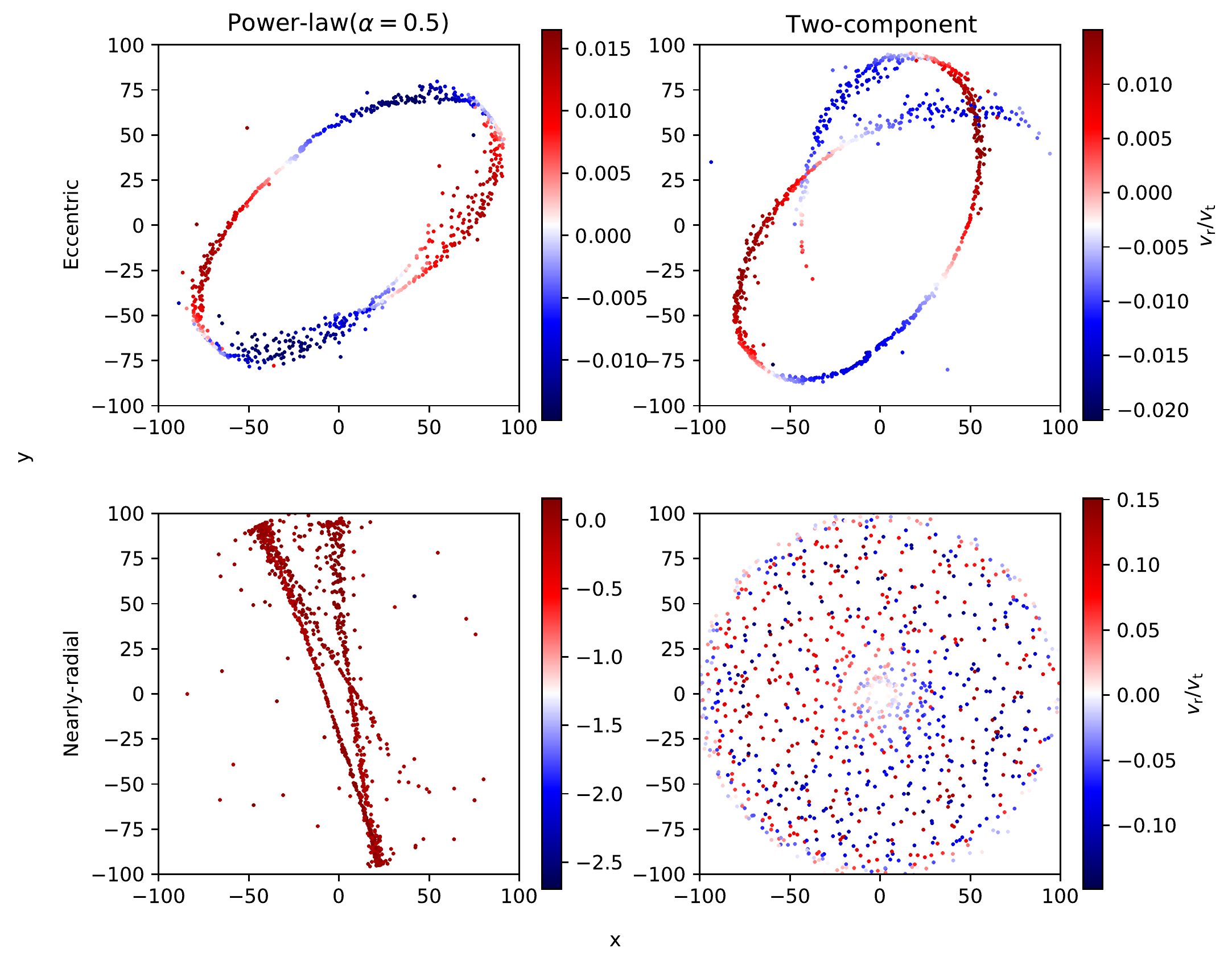}
    \caption{The positions of stars in the orbital plane ($x$-$y$) in the galactic frame. Colors represent the ratio between radial and tangential velocities of stars. The upper and lower panels show models with eccentric and nearly-radial orbits, respectively. The left and right columns show the power-law and two-component potentials, respectively. \label{fig:eccorbit}}
  \end{figure*}

  To explore the possibility of injecting the debris stars from the disrupted clusters to the 
  immediate neighborhood of SMBHs, we compare models with modestly and highly eccentric orbits as described 
  in \S\ref{sec:eccinit}. Figure~\ref{fig:eccorbit} show the spatial distribution of stars at about $92~\trhz$ 
  for four models with a static ($\rrho=0$), power-law ($\alpha=0.5$) and a static, two-component potential 
  ($\alpha=0.5$ and $\fbh(\Rg)=0.01$).

  For the modestly eccentric orbit, long tidal tails appear between $R \simeq 50-100~\rtz$ but no 
  cluster stars enter into the region with $R \lesssim \Rd \simeq 40~\rtz$ in both power-law
  and two-component (with SMBH) galactic potentials.  For the cluster with a highly-eccentric 
  (nearly-plunging) orbit, the outcome of peri-galacticon 
  passage is very different for these two types of galactic potentials.
  Under the power-law potential the cluster stars are distributed along narrow bridges and tails 
  along the original orbit of the star cluster.  Under the two-component (galaxy+SMBH) potential, 
  the disrupted stars widely spread out in all regions.  This feature indicates that the SMBH has 
  a strong scattering impact on the stars in the tidal debris of the disrupted star clusters 
  as they venture to its proximity.

\begin{figure}[ht!]
    \plotone{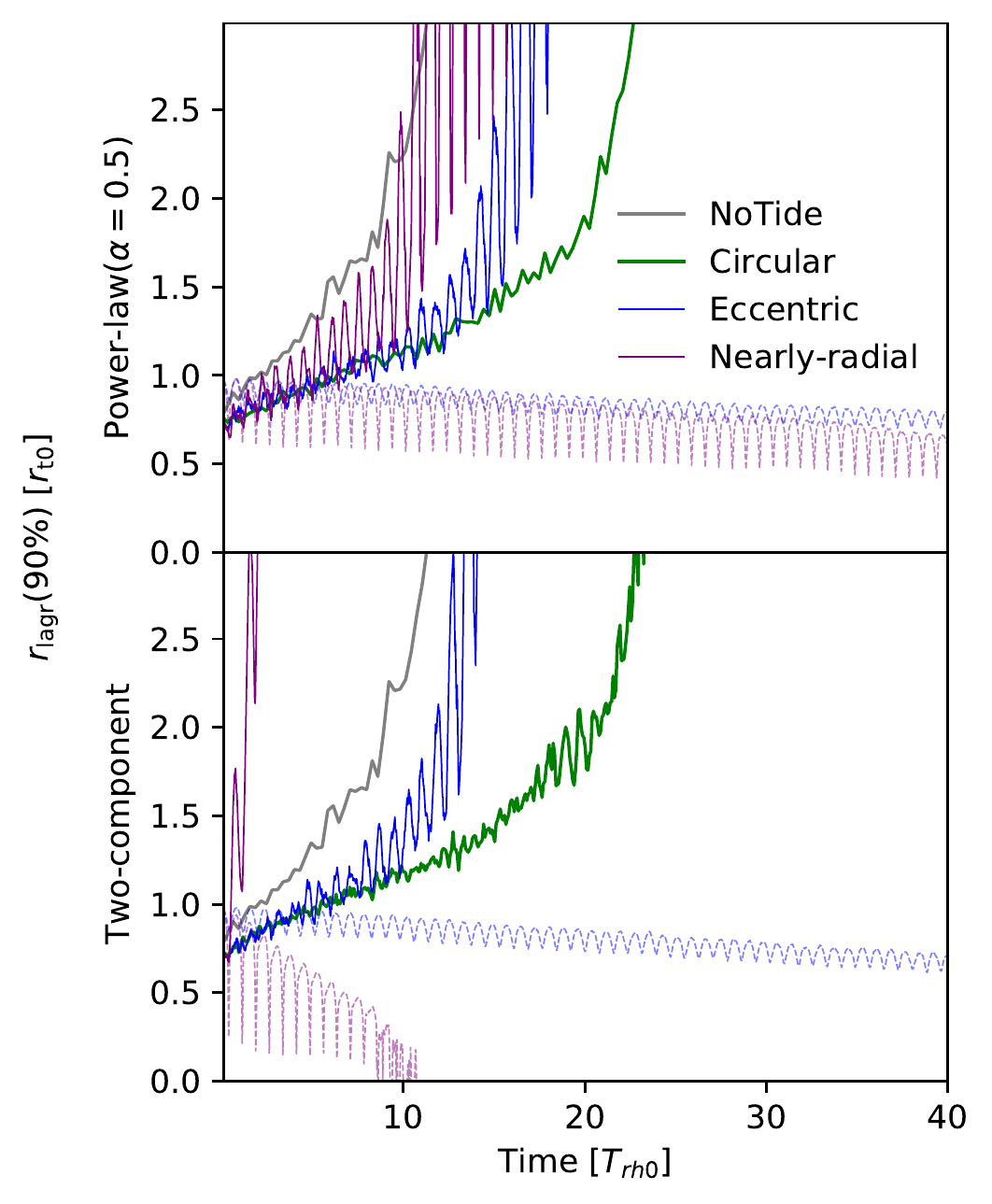}
    \caption{The evolution of  $\rl(90\%)$ (solid lines) and $\rt$ (dashed lines) for models with different orbits in a power-law ($\alpha=0.5$) and a static two-component potentials are compared. The reference model with no tidal field is shown as grey color. The models with a circular, an eccentric and a nearly-radial orbits are shown with different colors. \label{fig:ecclagr}}
  \end{figure}

  Figure~\ref{fig:ecclagr} shows the evolution of $\rl(90\%)$ and $\rt$, analogous to Figure~\ref{fig:kingpselagr}.
  For both modestly-eccentric and nearly-radial models under the power-law potential, the effect of tidal 
  compression before $\rl(90\%)<\rt$ remains apparent.
  The nearly-radial model has a rapid $\rh$ expansion due to a strong mass loss which also lead to a fast reduction
  of $\rt$.  Under the tidal influence of the SMBH, the cluster with a nearly-radial orbit quickly suffers severe tidal disruption.

  The ratio between the period of star clusters on a circular orbit at $\Rd=100~\rt$ and $\trhz$ is about 1.57.
  The modestly-eccentric clusters have shorter orbital periods.
  The frequent oscillation of $\rl$ and $\rt$ is due to the cluster's motion between apo- and peri-galactic passages.
  For both the modestly-eccentric and nearly-radial orbits, the tidal disruption occur smoothly over a few orbits. 

\begin{figure}[ht!]
    \plotone{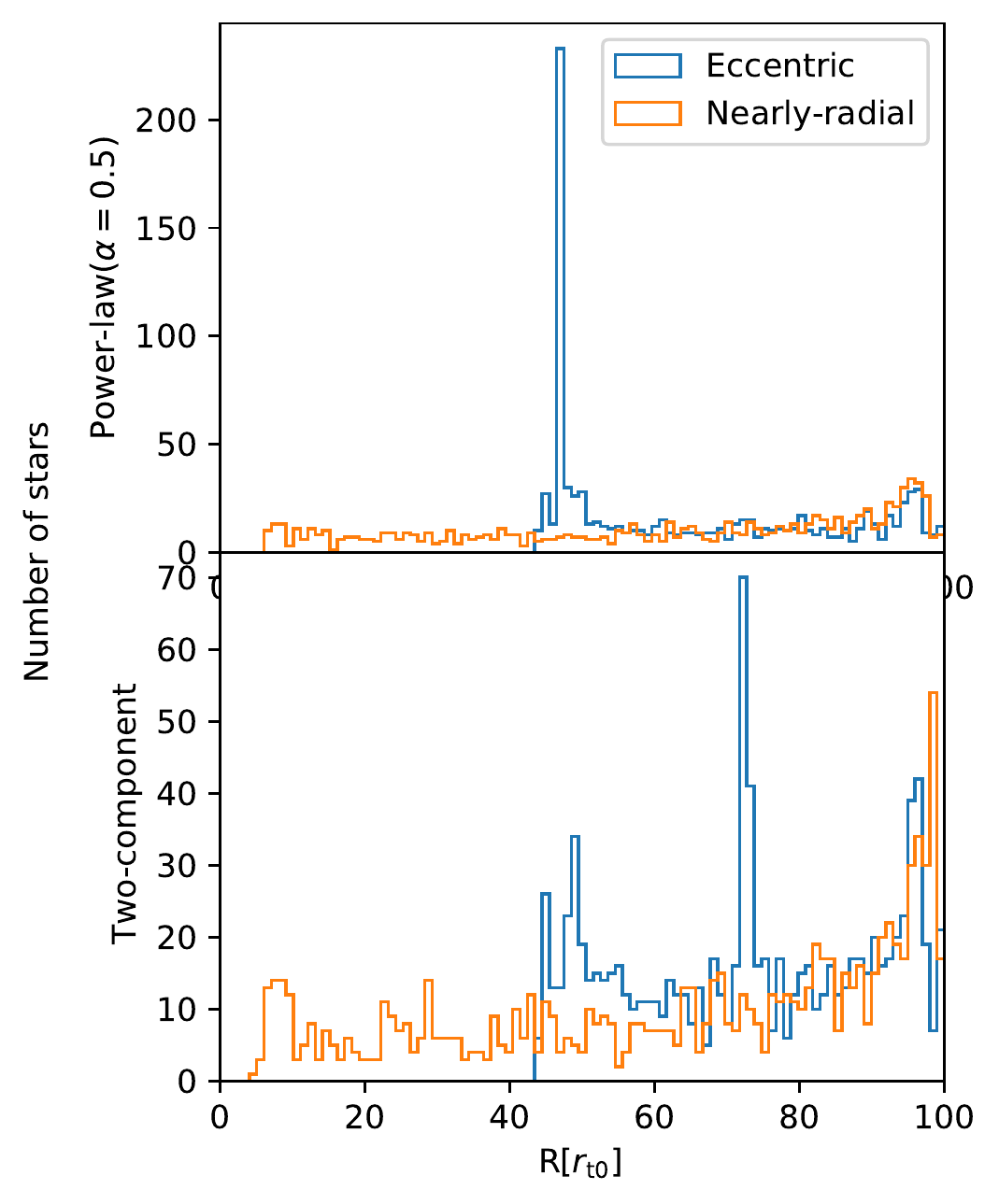}
    \caption{The radial distribution of stars referring to the galactic center for models with eccentric and radial orbits. The upper and lower panel show models with a power-law ($\alpha=0.5$) and a static two-component potentials respectively. \label{fig:eccnhist}}
  \end{figure}

  Figure~\ref{fig:eccnhist} shows the radial distribution of stars relative 
  to the galactic center  at $92~\trhz$. 
  The peak of the distribution indicates the location of star cluster. 
  There are no stars within $R \simeq 40~\rtz$ for the models with modestly-eccentric orbit.  But, the SMBH has a strong impact on the distribution of stars for the cluster with nearly-radial orbit.
  A significant fraction of stars can reach inside the tidal disruption distance ($\Rd \simeq 40~\rtz$).
  In contrast, in the case with gradually in-spiraling decay, dynamical friction is quenched
  when a star cluster is completely tidally disrupted near $\Rd$ so that only a few stars can reach 
  inside $\Rd$ (Fig. \ref{fig:kingpse1}). 

\begin{figure*}[ht!]
    \plotone{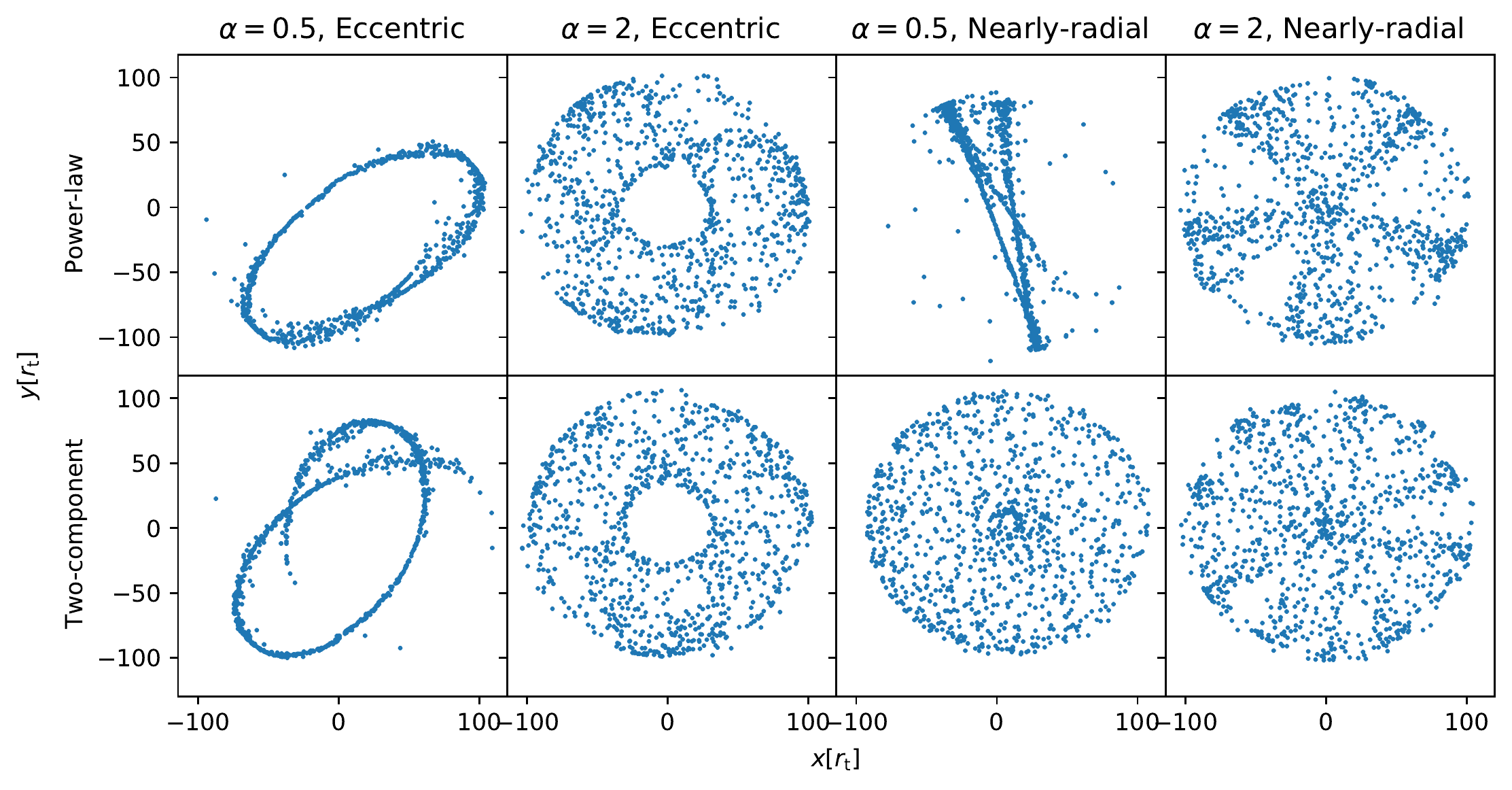}
    \caption{The $x$-$y$ distribution of stars referring to the cluster center at $92~\trhz$ for models with eccentric and radial orbits. The upper and lower panel show models with a power-law and a static two-component potentials respectively. For the power-law component, results of $\alpha=0.5$ and $2$ are compared for both eccentric and radial orbits. \label{fig:xyecc}}
  \end{figure*}

  Figure~\ref{fig:xyecc} shows the $x$-$y$ and  distribution of stars for models with eccentric and 
  radial orbits and different potentials at $92~\trhz$.
  For the single power-law potential, clusters still keep a narrow shape when $\alpha=0.5$ but
  their tidal debris spreads over much more extended region in the model with $\alpha=2$.
  This difference provides another evidence that tidal compression in galaxies with low $\alpha$ 
  density distribution delays the disruption of clusters.

\begin{figure*}[ht!]
    \plotone{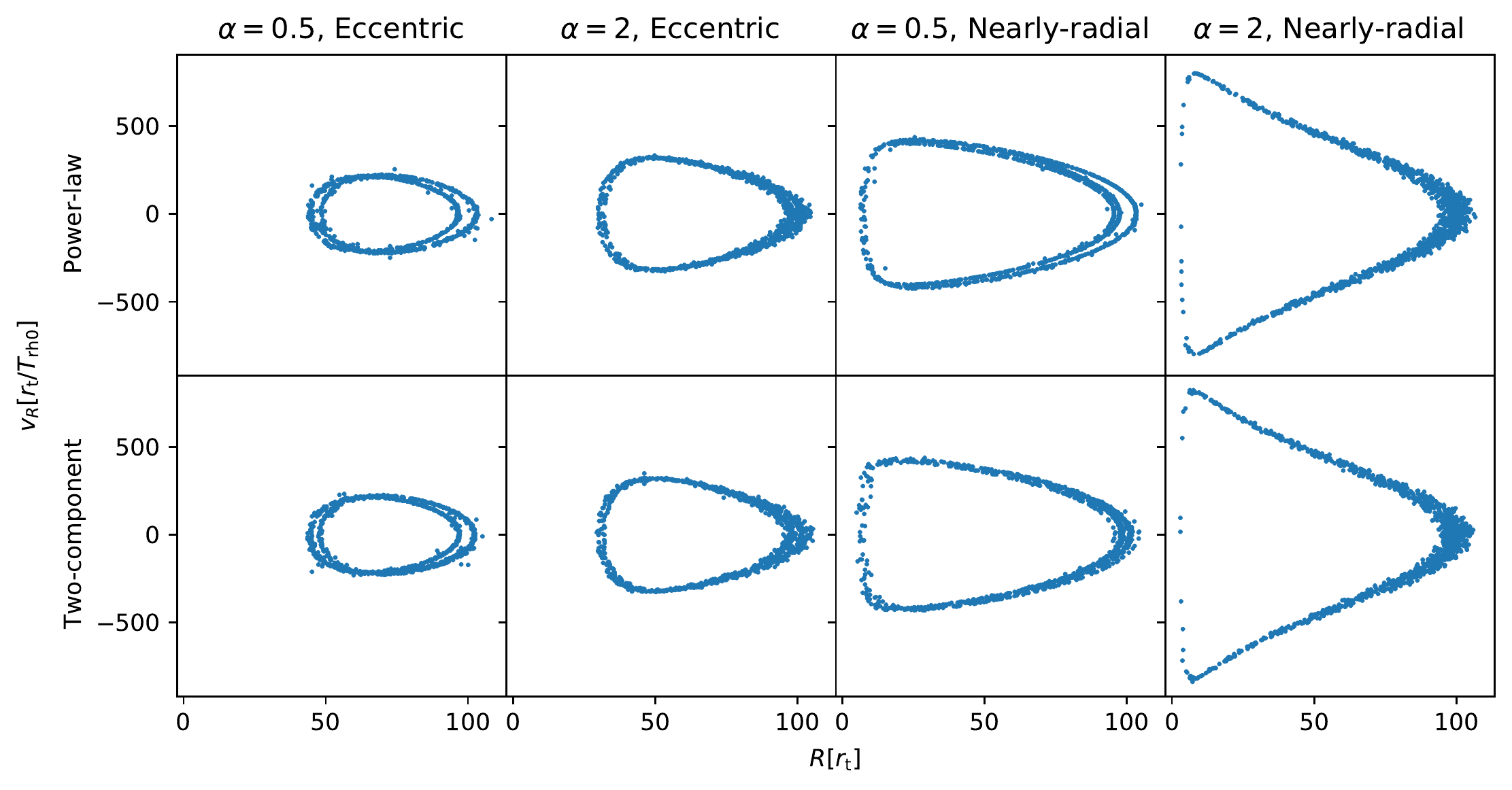}
    \caption{Similar to Figure~\ref{fig:xyecc}, but the radial velocity vs. $R$ of stars are shown.  \label{fig:vrrecc}}
  \end{figure*}

  Due to fast differential precession, the stars removed from the cluster quickly disperse to
  establish a nearly isotropic (in the cluster's original orbital plane) distribution.  
  Nevertheless, they retain the kinematic properties of their original host clusters.  
  Figure~\ref{fig:vrrecc} shows the $v_{\mathrm{R}}$-$R$ distribution for the same group of models. 
  All models show a clearly correlated patterns. Models with $\alpha=2$ show large $v_{\mathrm{R}}$.
  In comparison, more stars occupy the central region of the galaxy in models with $\alpha=0.5$ 
  (Fig. \ref{fig:xyecc}) where they are near their orbital peri-center with smaller $v_{\mathrm{R}}$
  (Fig. \ref{fig:vrrecc}). In principle, this distribution can be extracted from the observed radial 
  velocity proper motion of stars near the galactic center and be used to identify
  lost members of tidally disrupted clusters.

  \section{Summary and Discussions}
  \label{sec:summary}

  This work is primarily motivated by the possibility of populating the innermost 
  region of galaxies by inward migration of stellar clusters formed elsewhere. The main physical
  processes which may lead to such migration includes dynamical friction, mergers, and secular
  perturbation by galactic companions and satellites (see references of previous investigations
  in \S\ref{sec:intro}).  The main focus of this paper is whether clusters can reach the central 
  region. To highlight this outstanding issue, we apply our simulated results to the interpretation 
  of observational data.  \cite{do2020} found two populations of stars near (within $\sim 4$ pc in 
  projected distance from) the Galactic center, separated by different velocity dispersion and metallicity.
  We revisit the suggestion that these diverse stellar populations are the debris of disrupted stellar 
  cluster.  The mass contained in the NSC in this region is $M(R) \sim$ a few $10^7 M_\odot$
  in addition to that of the Sgr$^\star$ SMBH.  If the parent clusters have $\Msc \sim 1-2 \times 10^6 
  M_\odot$, $\rh \sim$ a few pc, and $r_{\rm cut} \sim {\mathcal O} (10 \rh$) (those of the most massive 
  globular cluster in the Galaxy), their conventional galacto-centric tidal disruption distance
  would be at least an order of magnitude larger than the location where diverse stellar populations
  were found (\S\ref{sec:homogenous} and \S\ref{sec:michie}). Even with the most-compact (and relative
  low-mass) cores of known globular clusters, their conventional tidal radius $\rt$ at $R \sim$ a few pc
  is $\lesssim$ their $\rc$ (Eqs. \ref{eq:rt} and \ref{eq:rcut}), i.e. they are expected to have disrupted at 
  larger galacto-centric distances. Note that the existence of a hypothetical IMBH at the center of an 
  in-spiral star cluster \citep{arcasedda2018} can provide a lower limit of $\rt$ depending on its mass ($M_{\mathrm{IMBH}}$). 
  For $M_{\mathrm{IMBH}} \simeq 10^4 M_\odot$, the corresponding $\rt\approx 0.3$~pc at $R=4$~pc from the Galactic 
  center, which is roughly the core radius of a dense star cluster.  Thus, an IMBH may help to prevent the 
  core from tidal disruption and bring it further towards the center of the Galaxy, albeit relatively
  small number of cluster stars may migrate to the last few pc from Sgr A$^\star$. 

  Our first attempt to bypass this migration barrier (imposed by the conventional, Roche-lobe-filling, necessary 
  tidal disruption condition)
  is to consider the possibility of tidal compression.
  In this work, we carry out a series $N$-body simulations of equal-mass star clusters under different 
  types of galactic potential, in order to determine the sufficient criterion for compressive
  tidal perturbation in 
  addition to the conventional necessary Roche-lobe-filling condition for tidal disruption of satellite
  stellar systems. We find that for a smooth power-law, galaxy-only, potential where $\alpha<0.913$, star clusters 
  with a circular orbit suffer tidal compression instead of tidal disruption (Figure~\ref{fig:homo}).
  The effect is independent of the density profiles of clusters (Figure~\ref{fig:kingw0ps}).

  In the case of two-component galaxy+SMBH potential, the boundary of tidal disruption and compression 
  (Figure~\ref{fig:kingpsbh}) depends on $\fbh$ (or equivalently $R$ or $\Rs$).  Analytic approximation 
  indicates that, with $\alpha=0.5$, for $\fbh(\Rg)<1/3$, there is a range of galacto-centric 
  distance from the SMBH smaller than the conventional Roche limit (the necessary condition), where 
  the tidal perturbation on the cluster is compressive.  But for $\fbh(\Rg)>1/3$, the
  sufficient criterion for tidal disruption is satisfied when the conventional, Roche-lobe-filling necessary 
  condition is met.
  Numerical simulations confirms this expectation.  In models with a fraction of the cluster stars  
  initially outside $\rt$ (Figure~\ref{fig:varyR}), these also undergo temporary tidal compression before they 
  are tidally removed from their host clusters. The observed density distribution of some galactic bulge 
  have $\alpha<1$ and we expect star clusters in these host environment to be tidally compressed in the 
  radial direction with respect to their centers.

  When a star cluster induces and endures dynamical friction, it sinks towards the galactic 
  center on an in-spiraling orbit.  With a representative prescription for the
  cluster's orbital decay in terms of an exponentially increases of galactic and SMBH's scaled masses 
  (\S \ref{sec:dfriction}), we find that tidal disruption still occurs (Figs.~\ref{fig:kingpselagr} and 
  \ref{fig:kingpselagr2}). Although the tidal compression is effective in the galacto-centric direction
  stars can escape along the tangential direction 
  (Fig.~\ref{fig:kingpse1orbit}). In general, star clusters cannot avoid tidal disruption near 
  the conventional tidal disruption distance $\Rd$ during tightly wrapped course of 
  their in-spiraling orbital decay.   Stars in the tidal debris carry similar specific orbital 
  energy and angular momentum, relative to the galactic center, as their parent clusters shortly prior to 
  their dispersal such that very few stars can venture inside $\Rd$ (Fig. \ref{fig:kingpse1}).

  In a follow-up attempt, we investigate excursion in the proximity of the galactic 
  center by clusters with modestly-eccentric and nearly-radial orbits. These orbits may be the 
  results of mergers of their host galaxies or be induced by secular perturbation of other 
  satellite galaxies. In our simulations, tidal compression is observed (Figure~\ref{fig:ecclagr}) 
  for both types of orbits.  Long tails of tidal debris also 
  appear (Figure~\ref{fig:eccorbit}).  During and after the disruption of a cluster 
  on a nearly-radial orbit, a significant fraction of the stars in the tidal debris retain $R \lesssim 40~\rtz$,
  including some with $R \sim 10~\rtz$(Fig. \ref{fig:eccnhist}). This outcome is different from the case of 
  an in-spiral cluster with a tightly-wrapped orbit, where the detached stars' closest distance $> 40~\rtz$
  (Fig. \ref{fig:kingpse1}).  In general, clusters with plunging orbits carry much less specific angular
  momentum around the galactic center and they venture to much smaller $R$'s.
  Although they eventually suffer tidal disruption, their stellar debris retain the clusters' 
  original kinematic properties (similar to comets' tails in the solar system) and the
  detached stars can reach much closer to the SMBH.

  Our $N$-body simulations provide supporting evidence to the hypothesis that the diverse stellar
  populations within a few pc's from the Galactic nuclei may have originated from both the in-situ star formation and the accretion of 
  globular clusters or dwarf galaxies \citep{feldmeier2014, tsatsi2017} and delivered to the
  nearest proximity of Sgr A$^\ast$ \citep{arcasedda2018} without the requirements of a hypothetical IMBH in or an exceptionally compact initial structure of the clusters.
  The SMBH in the Galaxy has the mass of $4.28\times 10^6 M_\odot$ \citep{Gillessen2017}.
  Based on the MWPotential2014 from \textsc{galpy} \citep{Bovy2015}, the enclosed Galactic mass within 4~pc is about $3.1\times 10^6 M_\odot$.
  With $\Msc \approx 2\times10^5 M_\odot$, and $\rc\approx 1$~pc, a typical globular cluster has 
  $\rt \approx 0.83$~pc for $\Mg \approx 7.38\times 10^6 M_\odot$. 
  It is possible for such clusters to survive tidal disruptions before reaching this distance 
  via either in-spiral induced by dynamical fraction  or on an initially plunging orbit. 
  For the in-spiral orbit, only the cluster core is left at 4~pc, and the halo is tidally stripped.
  \cite{arcasedda2015} argues that such low-mass globular clusters contribute little to the formation of the nuclear 
  star cluster. Indeed, the cluster core only contain a few hundred objects and most of them are stellar-mass black 
  holes. But with nearly-radial orbits, the stellar debris from a disrupted cluster may reach $R$ which is 4 
  times smaller than $\Rd (\sim 0.4 \Rg)$ (Fig. \ref{fig:eccnhist}).  It is therefore possible for 
  such orbit, the contribution from globular cluster to the nuclear cluster is somewhat enhanced.
  To identify this contribution, one possible way is to measure the orbit of stars and to identify 
  some $v_{\mathrm{R}}-R$ correlations (Fig. \ref{fig:vrrecc}), if any, within a few pc's from 
  the Galactic center.

  Our model is a highly simplified approximation to the more general processes of star-clusters' orbital
  evolution including the concurrent evolution of their host-galaxy's mass and potential, drag by and
  accretion of gas outside and inside the clusters, and the evolution of cluster stars. These additional
  processes are likely to affect the course and destiny of in-spiralling clusters during the early phases
  of galaxy formation and evolution.  Here we specifically 
  focus on the tidal compression effect from a group of power-law potential with $\alpha<1.0$. 
  We find that this effect cannot dramatically change the fate of star cluster, but it can somewhat 
  delay the disruption and 
  enable the stellar debris to settle in the proximity of the Galactic center. 

  \begin{acknowledgments}

    L.W. thanks the support from the one-hundred-talent project of Sun Yat-sen University, the Fundamental Research Funds for the Central Universities (22hytd09), Sun Yat-sen University and the National Natural Science Foundation of China through grant 12073090 and 12233013. 
    We thank P. Ivanov, T. Do, and Z. Chen for useful conversation.  We also thank an anonymous referee for 
    thorough reading and detailed suggestions which have helped us to improve the presentation.  

  \end{acknowledgments}

  %

  \vspace{5mm}


  \software{\textsc{petar} \citep{Wang2020b},
    \textsc{sdar} \citep{Wang2020a},
    \textsc{fdps} \citep{Iwasawa2016,Iwasawa2020},
    \textsc{galpy} \citep{Bovy2015},
    \textsc{mcluster} \citep{Kuepper2011}
  }




  \appendix

  \section{Dynamical friction of a point mass}
  \label{sec:appendixa}
  We approximate the dynamical friction by increasing the galactic mass with the controlling parameter of $\rrho$.
  Since we focus on how tidal compression works for an in-spiral orbit.
  The major impact is the evolution of $\rt$ as $R$ decreases.
  The speed of in-spiral as controlled by $\rrho$ does not suppress tidal compression but only affect the timescale.
  Thus, discrepancies in the in-spiral speed between our idealized approximation 
  and the conventional formulae does not change our general conclusion.

  Just for comparison, we integrate a point mass orbit in a power-law potential with $\alpha=0.5$. The initial condition of the point mass follows that of the center-of-the-mass of the star cluster shown in Figure~\ref{fig:kingpse1}. 
  In addition to the gravitational force from the galaxy, we include the friction force based on 
  the conventional \citet{Chandrasekhar1943} dynamical-friction formula \citep{binney2008}:
  \begin{equation}
    \frac {\mathrm{d} \mathbf{v}}{\mathrm{d} t} = - \frac{4 \pi^2 G^2 M \rhog \ln{\Lambda}}{ v^3} \left[ {\rm erf} 
      (X) - \frac{2X}{\sqrt{\pi}} e^{-X^2} \right] \mathbf{v}
    \label{eq:fric}
  \end{equation}
  where $\mathbf{v}$ is velocity vector of the point mass and $X \equiv v/\left(\sqrt(2) \sigma \right)$.
  The one-dimensional velocity dispersion is evaluated by assuming a virial equilibrium as $\sigma = \sqrt{\Psi_{\mathrm{g}}(R)/3}$.
  We assume the velocity distribution of field stars is Maxwellian with an isotropic dispersion of $\sigma$.
  The Coulomb logarithm ($\ln{\Lambda}$) affects the timescale of dynamical friction.
  We simply adopt it to be 10.

  As star cluster loses mass, the efficiency of dynamical friction weakens.
  To approximate the mass loss, we assumes the mass of the point linearly depends on $R$:
  \begin{equation}
    m_{\mathrm{p}} = m_{\mathrm{p0}} \frac{R-0.4 \Rg}{0.6 \Rg}.
  \end{equation}
  The factor 0.4 represents the transition shown in Figure~\ref{fig:kingpse1}, where the 
  cluster lose equilibrium and reach the phase of fast disruption.

\begin{figure}[ht!]
    \plotone{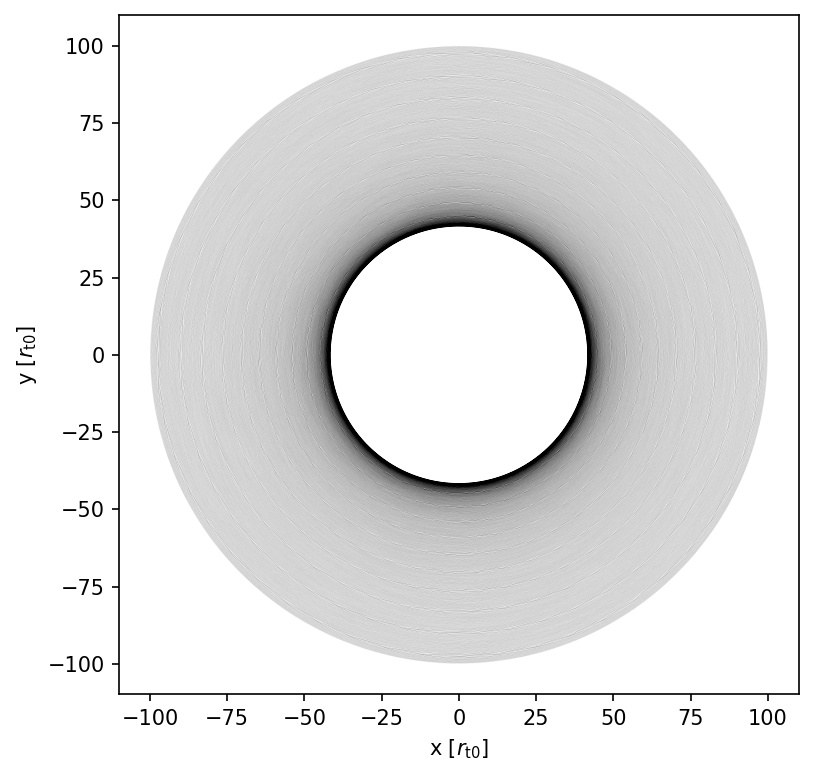}
    \caption{Orbit of a point mass under dynamical friction in the power-law potential with $\alpha=0.5$. \label{fig:pmorbit}}
  \end{figure}

  With this setup, we integrate the orbit of the point mass until it reaches the minimum $R=0.4\Rg$.
  The orbital trajectory is shown in Figure~\ref{fig:pmorbit}.
  Similar to Figure~\ref{fig:kingpse1} (where the idealized prescription for dynamical friction is applied), 
  the radial migration rate decrease when $R$ is close to $0.4~\Rg$, as the efficiency of dynamical friction weakens.
  The migration rate can vary with different value of $\ln{\Lambda}$ and $\alpha$.  Although
  our idealized prescription of dynamical friction does not exactly reproduce cluster's exact orbital path 
  computed with the conventional dynamical-friction for this set of model parameters, it match the general 
  trend.

  \bibliography{reference}{}
  \bibliographystyle{aasjournal}


\end{CJK*}
\end{document}